\documentclass[showpacs,preprintnumbers]{revtex4}
\usepackage{amsfonts}
\usepackage{amssymb}
\usepackage{amsmath}
\usepackage{epsfig}
\usepackage{array}
\allowdisplaybreaks
\begin{document}
\title{Dynamics of self-gravitating systems in non-linearly magnetized chameleonic Brans-Dicke gravity}

\author{Z. Yousaf}
\email{zeeshan.math@pu.edu.pk} \affiliation{Department of
Mathematics, University of the Punjab, Quaid-i-Azam Campus,
Lahore-54590, Pakistan.}

\author{Kazuharu Bamba}
\email{bamba@sss.fukushima-u.ac.jp} \affiliation{Faculty of
Symbiotic Systems Science, Fukushima University, Fukushima 960-1296,
Japan}

\author{M. Z. Bhatti}
\email{mzaeem.math@pu.edu.pk} \affiliation{Department of
Mathematics, University of the Punjab, Quaid-i-Azam Campus,
Lahore-54590, Pakistan.}

\author{S. Rehman}
\email{sanarehman718@gmail.com} \affiliation{Department of
Mathematics, University of the Punjab, Quaid-i-Azam Campus,
Lahore-54590, Pakistan.}


\begin{abstract}
We study the effects of magnetic fields of non-linear
electrodynamics in chameleonic Brans-Dicke theory under the
existence of anisotropic spherical fluid. In particular, we explore
dissipative and non-dissipative self-gravitating systems in the
quasi-homologous regime with the minimal complexity constraint. As a
result, under the aforementioned circumstances, several analytic
solutions are found. Furthermore, by analyzing the dynamics of a
dissipative fluid, it is demonstrated that a void covering the
center can satisfy the Darmois criteria. The temperature of the self
gravitating systems is also investigated.
\end{abstract}
\maketitle

\section{Introduction}

The mechanism through which the individual elements of a large body
are kept together by the collective gravity of the structure as a
whole is termed as self-gravitation. One of the few observable cases
in which general relativity (GR) is likely to play an essential role
is the gravitational collapse of large bodies such as black hole
(BH) and neutron star. As a result, a precise explanation of
gravitational collapse and the evolution of the structure of the
universe was required. Stellar objects under a range of different
situations remained one of the most intriguing challenges in GR
\cite{herrera2009structure}. There was an enormous appeal towards
these issues by the relativists community. Beginning with
Oppenheimer and Snyder's fundamental articles
\cite{oppenheimer1939continued}, an extensive list of works by
Oppenheimer and Volkoff \cite{oppenheimer1939massive} has been
offered to investigate the models of gravitating systems and stellar
objects developing at that time. Gravitational waves are an advanced
application of gravitational collapse which has been under study for
the last four decades \cite{fryer2011gravitational}. Different
space-based and ground-based interferometric detectors
\cite{sasaki2003analytic,jafry1994lisa} were used for the detection
of gravitational waves linked with gravitational collapse. As we are
interested in the gravitational collapse of self-gravitating
systems, the formulation of large galaxies may lead, one to believe
that any vast number of objects in such systems were held together
by their mutual gravitational attraction. In general relativistic
models, the behavior of a fluid distribution after it departs from
the equilibrium state was studied when non-zero radial forces
appeared with opposite signs inside the self-gravitating system. The
anisotropy of the self-gravitating system was demonstrated to be
closely connected to this impact \cite{herrera1997local}.

The study of self-gravitating systems was extended by introducing
the concept of complexity. Complexity is a physical term strongly
entwined with the fundamental characteristics of the system. The
complexity factor of a system can be created by various physical
parameters such as viscosity, density inhomogeneity, etc. A system
with regular energy density and isotropic pressure might be
considered as the fundamental system with a null (zero) complexity
factor. The complexity factor is a scalar function used to calculate
the fluid distribution complexity level. The complexity of the
self-gravitating systems was further extended for the time-dependent
system. While considering such systems, two challenges were faced,
i.e., generalization of the the case for time-dependent dissipative
models and calculation of the complexity of evolution patterns
\cite{herrera2018new}. Herrera \emph{et al.}
\cite{herrera2018definition} offered a formal study of static
spherically symmetric systems by considering the complexity factor
of fluid configuration. Dynamical variables are the most crucial
components playing role in the evolution of self-gravitating
systems. Herrera \cite{herrera2009structure} discovered the
orthogonal splitting of curvature tensor by using five scalar
functions known as structure scalars to study the configuration as
well as the pattern of the evolution of self-gravitating fluid
distribution. These scalars are demonstrated to be linked directly
to the fundamental characteristics of the fluid configuration such
as local pressure anisotropy, energy density, inhomogeneity in
energy density, heat flow, etc. Herrera
\cite{herrera2019complexity1, herrera2019complexity,
herrera2020stability} determined the complexity of the fluid
distribution in various geometric configurations such as Bondi
metric, axially spherically symmetric systems to deeply investigate
the concept of this factor.

{Abbas and Nazar \cite{abbas2018complexity} acknowledged the
complexity factor for an anisotropic static matter distribution
using non-minimal metric $f(R)$ gravity. Utilizing certain
observational information from well-known compact bodies, they
attained modified field equations, mass function, the physical
characteristics of $f(R)$ model, and scalar functions. Also, various
solutions are obtained for the presented model. Yousaf \emph{et al.}
\cite{yousaf2020new,yousaf2022structure} established complexity factor for cylindrically
symmetric fluid distribution in $f(R, T, R_{\mu\nu}T^{\mu\nu})$
gravity. They have analyzed the static irrotational cylindrically
symmetric anisotropic fluid to formulate the gravitational field
equations, structure scalars, etc. They determined the solutions for
the modified gravity and compared them with GR.}

The homologous condition was considered to be the easiest mode of
evolution but this condition was relaxed further to get more
feasible solutions known as quasi-homologous $(Q_h)$ regime
\cite{herrera2020quasi}. In the context of GR, dynamical as well as
analytical solutions associated with hyperbolic symmetry satisfying
minimal complexity condition are presented. Such types of solutions
evolve in the $Q_h$ condition. In the field of modified gravity,
Yousaf \textit{et al.} \cite{bhatti2021role, yousaf2016influence}
discovered the same variables. The function of these scalar
quantities in the formation of inhomogeneity of energy density in
case of self-gravitating anisotropic fluid distribution was invested
by Herrera \cite{herrera2018new}. The dissipative and
non-dissipative systems for fluid distribution are studied
separately. In the case of dissipative fluid, a geodesic fluid is
created by imposing the vanishing complexity factor ($Y_{TF}$) and
$Q_h$ condition. If one consider a non-dissipative fluid
(time-dependent), the fluid is also shear-free endowed with
isotropic pressure and homogeneous energy density. In the general
dissipative case, fluid is shearing resulting in the formation of a
family of solutions. Variations in isotropy and deviations in local
anisotropy can be induced by a large range of physical processes
often found in celestial objects. The presence of dissipation,
shearing as well as energy density inhomogeneities has recently been
demonstrated to make the isotropic condition unstable. There has
been a very attractive results on the study of compact objects and
dark energy problem \cite{astashenok2014maximal,
astashenok2013further, Nojiri:2006ri, nojiri2017modified}.

Gravitational collapse is a well-known example of a strong-field
system. Many researchers have been interested in studying it in
modified theories \cite{sharif2013dynamical, harada1997scalar}.
Penrose \cite{penrose1979global} extensively detailed the
implications of the conformal tensor in the existence as well as the
evolution of isotropic fluid distribution. He also described the
stability of such spherically symmetric systems. The importance of
anisotropic fluid, as well as shearing and shear-free systems in
self-gravitating systems, is well acknowledged by different
researchers. By using the definition of complexity for a spherically
symmetric static model, anisotropic star models have been formed.
These models fulfilled the minimal complexity condition to enhance
the information by enclosing Einstein field equations. By offering
new study avenues, GR has provided new approaches for exploring the
universe. In the last century, GR has made significant progress in
the study of relativistic models as well as in the evolution of the
universe. The most intriguing problem in the recent study of
cosmology is understanding the process that led to a recent phase of
rapid expansion of the universe. The evidence that we are living in
an expanding universe is overwhelming
\cite{pietrobon2006integrated}. The particular nature of the
cosmological term in GR which causes the expansion in the universe
is still unknown. A significant number of modern theories are
proposed to understand the enigma of the accelerated expansion of
the universe by analyzing the Einstein-Hilbert action. Numerous
experimental instruments including Supernova, WMPA, X-ray emission
from galaxy stars, etc. \cite{riess1998observational,
caldwell2004cosmic} indicate the accelerating expansion as strong
radiation. This fast expansion of the universe is driven by a
peculiar type of matter that dominates the universe structure. All
theories of gravity are compatible with GR in weak-field systems but
they may diverge significantly from GR in strong-field systems.
Modified theories in strong-field systems are thought to lead to an
appropriate theory of gravity.

For various reasons, GR has generated several alternative theories,
one of which is scalar-tensor field theory. A non-minimal coupling
expression is used to bring the scalar field under consideration in
a non-trivial way \cite{dunn1974scalar}. The scalar-tensor field
theory was originally determined by Jordan
\cite{jordan1948funfdimensionale}, who started to implant the curved
4-dimensional manifold into flat 5-dimensional space-time. Jordan's
attempt was taken over by Brans and Dicke \cite{brans1961mach}.
Brans Dicke (BD) theory is a scalar-tensor field with many
intriguing features, such as a gravitational constant (
time-dependent), direct geometry-scalar field interaction, and
compliance with many physical laws. The dimensionless quantity
$w_{BD}$, being a constant coupling quantity, is the only parameter
of this theory. BD theory has conformal invariance which is similar
to string theory at high energies. BD scalar $\phi$ has asymptotic
behavior due to which BD reduces to GR as $w_{BD}\rightarrow\infty$
\cite{faraoni1998omega}. The astrophysical limitations derived from
solar-system observations predicted the coupling strength $w_{BD}$
to fulfill the condition $w_{BD}\geq40,000$ only for scalar-field
$\phi$ being massless \cite{bertotti2003test,
sharif2013thermodynamics}. It also gives evidence for a variety of
cosmic concerns, such as cosmic acceleration, the universe's late
behavior, and the inflation problem, etc.
\cite{sharif2012anisotropic, banerjee2001cosmic}. {Joudaki \emph{et al.} \cite{joudaki2022testing} presented a
comprehensive investigation of a particular modified theory in BD
gravity starting with a numerical and analytical depiction of the
linear perturbations and background expansion to the cosmological
constraints, large-scale bodies, etc. They modeled the non-linear
adjustments to the fluid power spectrum. Amani
\cite{amani2022energy} considered a modified BD theory in four
dimensions. He discussed the energy constraints as well as bounds in
the modified gravity using imperfect matter configuration and
checked whether the constraints provide a viable explanation of the
cosmos in the present era.}

In GR, a new method is being studied, in which the scalar field is
permitted to interact non-minimally with matter rather than
geometry, and this interaction is established via an interference
factor in the action integral \cite{khoury2004chameleon,
das2008brans}. This sort of scalar field is called the chameleonic
field. Various fascinating possibilities related to the chameleonic
field have been recently examined \cite{das2006superacceleration}.
This chameleon field has also recently been demonstrated to allow a
remarkably seamless transition from a decelerated to an accelerated
phase of universe expansion. This field has also non-minimal
interaction with matter. At this point, it is worth noticing that
efforts are made to construct models in which DE and dark matter do
not conserve themselves independently, but instead they interact
\cite{zimdahl2004statefinder}. The work on Chameleonic Brans-Dicke
(CBD) theory was inspired by Clifton and Barrow's
\cite{clifton2006decaying} in which they investigated the behavior
of the isotropic cosmological system in the early and late time
limitations in this framework. The current model may agree well with
the observed constraints even for very large values of
$w_{CBD}\sim10^4$. As a result, this type of general interaction has
properties that might answer cosmological difficulties as well as
the study of observations on solar systems. Different cosmological
applications in BD theory with the chameleonic scalar field for
holographic dark energy (HDE) are studied. The holographic dark
energy and age-graphic dark energy (ADE) theories have lately gained
a lot of attention among many scenarios to explain the acceleration
of the universe expansion in CBD theory \cite{wang2017holographic,
sheykhi2011interacting, setare2010holographic}.

In cosmological models, the existence of initial singularity (also
known as the Big Bang) is another fascinating topic for scientists
to investigate. To solve this problem, numerous models have been
proposed, including the cosmological constant
\cite{de1917relativity}, interactions between fields in Lagrangian
density, non-linear electrodynamics (NLED)
\cite{bandyopadhyay2012study}, and non-equilibrium characteristics
in thermodynamics \cite{murphy1973big}, etc. Primordial singularity
is an unfathomable condition characterized by infinite density and
curvature, in which all the physical, as well as mathematical
perceptions, are meaningless. The introduction of NLED  can give a
feasible solution to the fundamental difficulties of the universe
model. It generalizes Maxwell's field theory for strong-field
regimes. In Maxwell's electrodynamics, action is quadratic in
electromagnetic field tensor ($F_{\mu\nu}$), and hence Maxwell's
field equations are linear. However, Maxwell's theory may be
generalized to a non-linear theory due to the non-linear corrections
of the higher order of $F_{\mu\nu}$. Depending on the universal
model, the form and nature of these adjustments may change. Born and
Infeld \cite{born1934foundations} built the first non-linear
modification which proved to be a fascinating theory in many ways.
In 1935, Heisenberg \cite{dunne2012heisenberg} and Euler evaluated
an effective action expressing non-linear corrections to Maxwell's
field theory. Since then, different models of non-linear
electrodynamics have been extensively investigated such as string
theory, FRW universe, etc.

In early work \cite{novello2007cosmological}, it is shown that the
addition of non-linear variables in electrodynamics can successfully
depict the recent cosmic phase without any initial singularity. A
link between Einstein's field equations and BH electrodynamics was
conjectured after the thermodynamical discussion of BH in 1930. In
this regard, Jacobson \cite{jacobson1995thermodynamics} was the
pathfinder for such a relationship. For the Rindler model, he
established this connection by using the first law of thermodynamics
by making the entropy and horizon area proportional to one another.
In GR and extended theories, the validity of GSLT has become a
contentious issue. According to this law, the sum of the entire
matter entropy and horizon entropy encompassed by that horizon must
be time-dependent and non-decreasing. De Lorenci $et~al.$
\cite{de2002nonlinear} suggested a suitable approach for avoiding
the cosmic singularity via a non-linear modification with an
electromagnetic field. Exact solutions of Einstein's field equations
(EFEs) connected to NLED might reveal the importance of non-linear
terms in powerful gravitational and magnetic fields. Different
reasons can be used to justify the lagrangian and the
electrodynamics associated with the Maxwell theory. By assuming the
presence of symmetries such as Lorentz invariance, gauge invariance
\cite{sorokin2021introductory}, etc., non-linear components can be
introduced to the classical Maxwell lagrangian. It is worthy to note
that nonlinear corrections might be crucial in avoiding the black
hole singularity. Setare and Shafei \cite{setare2006holographic}
studied the apparent horizon in terms of the first and second law of
thermodynamics. Setare \cite{setare2007interacting} addressed the
validity of the generalized second law of thermodynamics in a closed
universe bounded by the event horizon using the interacting HDE. He
discovered that the second law is obeyed for a unique range of the
deceleration factor. GSLT has been analyzed in various modified
gravities such as scalar-tensor field theory, $f(R)$ cosmology,
\cite{dehghani2006thermodynamics,bandyopadhyay2011study} etc.
Recently, the feasibility of GSLT is studied in Chameleonic
cosmology \cite{farajollahi2011generalized}.

It is suggested \cite{novello2004nonlinear} that the modified terms
with positive power provide a convincing explanation of the
primordial singularity and inflation phase whereas the modified
terms with negative power can interpret the universe's late time
expansion. In our work, we employed the Euler-Heisenberg lagrangian
(including quadratic and linear terms only) in which the quadratic
parameters provide the bouncing era that precludes the primordial
singularity and linear parameters determine the radiation-dominated
era. It has been proposed that by utilizing the strong field limit,
the curvature reaches its minimal value, implying the universe's
smallest volume. Furthermore, if the squared terms of electric as
well as magnetic field are equal, then the universe has an early
cosmic singularity and the bouncing potential is avoided
\cite{garcia2009dynamics}. As a result, the Euler-Heisenberg
lagrangian is effective in both the early and middle ages. Since
scalar-tensor field theories are significant in describing dark
energy and its implications, CBD gravity is considered a promising
choice for characterizing late-time cosmic dynamics. As a result, we
have examined the action integral, a combination of scalar field
Lagrangian density and Euler-Heisenberg Lagrangian density, to
create a comprehensive model that can represent the primordial
cosmic expansion to the late-time expansion. We are going to study
the effects of non-linear electrodynamics in CBD gravity.

{A central idea in cosmology is the intriguing fact that the cosmos
is expanding at an accelerated rate. The latter is the most
prominent and tricky problem in astrophysics and cosmology that
creates an imbalance in the gravitational field equations.
Chakrabarti \emph{et al.} \cite{chakrabarti2022screening} presented
a method by which the scalar field in BD theory might escape local
astronomical constraints and serve as the primary source of
late-time acceleration of the cosmos. The scalar field acquires an
effective mass depending on density as the chameleonic field. They
analyzed the system's feasibility in light of solar system
constraints, the Equivalence Principle, etc. Finally, they concluded
that the cosmological constraints infer the presence of moderate
matter-scalar interaction in BD theory. The quasi-homologous
constraint is the simplest evolution pattern of self-gravitating
fluid distribution. Yousaf \emph{et al.} \cite{yousaf2022selfzna}
determined the $Q_{h}$ condition in the presence of a chameleonic
field in the BD framework. They evaluated the modified field
equations, mass function, scalar functions, complexity factor, etc.
They deduced that the complexity of the system increases due to the
effective mass of the scalar field by finding various solutions for
various dissipative as well as non-dissipative models.}

The goal of this paper is to provide a comprehensive overview of the
non-linear electrodynamics in the CDB theory and its physical
effects in energy-momentum tensor. The influence of NLED on the CBD
theory with matter contents such as anisotropic fluid, magnetic
field, and CBD field DE will be investigated in this paper. The
format of the paper is as follows. The magnetized CBD formulation in
the presence of anisotropic fluid and NLED, as well as the mass
function and the orthogonal splitting of curvature tensor in the
presence of non-linear electromagnetic field terms, will be
discussed in the next section. In Sec. III, junction conditions for
both metrics, i.e., inner and outer boundary, are examined for NLED
in magnetized CBD theory. Sec. IV depicts the $Q_h$ constraint for
the evolution of the system. In Sec. V, we seek further kinematical
restrictions to construct galactic models in the presence of
magnetic field. In Sec. VI, transport equation for evaluating the
temperature is determined. Sec. VII involves exact analytical
solutions for dissipative and non-dissipative models in magnetized
gravity. Finally, we review the findings from the previous sections.

\section{Formulation of Magnetized Chameleonic Brans-Dicke theory}

In this section, we are going to formulate the CBD theory in the
presence of non-linear electromagnetic field. For magnetized CBD
theory, the principle of least action can be written as
\cite{sharif2013thermodynamics}
\begin{equation}\label{a1}
S=\sqrt{-g}\int\bigg(L^{\phi}+{f(\phi)}(L^{m}+L^{ME})\bigg){d^{4}x},
\end{equation}
where {$L_{\phi}$} is lagrangian  density of $\phi$ i.e., scalar
field including non-minimal coupling with geometry and $f(\phi)$ is
the interaction of scalar-matter fields. One can write
\begin{equation}\label{a2}
L^{\phi}={\phi}R-U(\phi)-\frac{w_{CBD}}{\phi}\nabla^{\alpha}\phi\nabla_{\alpha}\phi,
\end{equation}
where $\phi$ is the CBD scalar-field which is the inverse of
Newton's constant, $R$ is the Ricci scalar, $w_{CBD}$ is the generic
dimensionless parameter of CBD theory and $U$ represents the CBD
field potential. In Eq. \eqref{a1}, $L^{m}$ represents the matter
distribution for locally anisotropic fluid with density $\rho^{m}$
and pressure $P^{m}$ \cite{herrera2020quasi}. We take matter
distribution as follows
\begin{equation}\label{a3}
T^{m}_{\mu\nu}=(\rho^{m}+P^{m}_{\bot})v_{\mu}v_{\nu}+(P^{m}_{r}-P^{m}_{\bot})\chi_{\mu}\chi_{\nu}+P^{m}_{\bot}g_{{\mu}{\nu}}+q_{\mu}v_{\nu}+q_{\nu}v_{\mu},
\end{equation}
where $\rho, v^{\mu}, q^{\mu}, \chi^{\mu}, P^{m}_{r}, P^{m}_{\bot}$
are energy density, four velocity, heat flux, a unit four-vector
(along the radial direction), radial pressure and tangential
pressure, respectively. The four-velocity, heat flux, and unit
four-vector are calculated as $v^{\mu}=\frac{1}{A}\delta^{\mu}_{0},~
q^{\mu}=q\frac{1}{B}\delta^{\mu}_{1} ,~
\chi^{\mu}=\frac{1}{B}\delta^{\mu}_{1}$. These quantities satisfy
\begin{equation}\nonumber
v^{\mu}v_{\mu}= -1,\quad\quad
v^{\mu}q_{\mu}=0,\quad\quad\chi^{\mu}v_{\mu}=0,
\quad\quad\chi^{\mu}\chi_{\mu}=-1,
\end{equation}
where under co-moving reference frame. The canonical form of Eq.
\eqref{a3} can be written as
\begin{equation}\label{a4}
T^{m}_{\mu\nu}={\rho^{m}}v_{\mu}v_{\nu}+\Pi_{{\mu}{\nu}}+P^{m}
h_{{\mu}{\nu}}+q(v_{\mu}\chi_{\nu}+\chi_{\nu}v_{\mu}),
\end{equation}
where
\begin{equation}\nonumber
P^{m}=\frac{2P^{m}_{\bot} + P^{m}_{r}}{3} , \quad\quad
h_{{\mu}{\nu}}=g_{{\mu}{\nu}}+v_{\mu}v_{\nu},
\end{equation}
\begin{equation}\nonumber
\Pi_{{\mu}{\nu}}=\Pi\bigg(\chi_{\mu}\chi_{\nu}-{\frac{1}{3}}h_{{\mu}{\nu}}\bigg),
\quad\quad \Pi=-\bigg(P^{m}_{\bot} - P^{m}_{r}\bigg).
\end{equation}
Moreover, $L^{ME}$ represents the lagrangian-density of non-linear
electromagnetic field. The lagrangian density of Maxwell's
electrodynamics can be written as
\begin{equation}\label{a5}
L=-\frac{1}{4 \mu_{0}}F_{\mu\nu}F^{\mu\nu},
\end{equation}
where $F_{\mu\nu}$ is the electromagnetic field tensor with
$F=F_{\mu\nu}F^{\mu\nu}$ and $\mu_{0}$ is the magnetic permeability.
The canonical stress-energy tensor can be written as
\begin{equation}\label{a6}
T_{\mu\nu}=\frac{1}{\mu_{0}}\bigg(F_{\mu\beta}F^{\beta}_{\nu}+\frac{1}{4}Fg_{\mu\nu}\bigg).
\end{equation}
The electromagnetic fields can be generated by the averaging
procedure of electric and magnetic fields \cite{de2002nonlinear}
which is given as
\begin{align}\nonumber
<E_i>=0,\quad <E_iE_j>=-\frac{1}{3}E^2g_{ij},\quad <B_i>=0,
\\\label{a7} <B_iB_j>=-\frac{1}{3}B^2g_{ij},\quad <E_iB_j>=0.
\end{align}
So from Eq. \eqref{a7}, we can get
\begin{equation}\label{a8}
<F_{\mu\beta}F^{\beta}_{\nu}>=\frac{2}{3}\bigg(\epsilon_{0}E^2+\frac{B^2}{\mu_{0}}\bigg)
+\frac{1}{3}\bigg(\epsilon_{0}E^2-\frac{B^2}{\mu_{0}}\bigg).
\end{equation}
Now, by comparing with the average value of stress energy tensor
\eqref{a4}, the energy density and pressure have the forms
\begin{align}\label{a9}
\rho^{m}=\frac{1}{2}\bigg(\epsilon_{0}E^2+\frac{B^2}{\mu_{0}}\bigg),
\quad P^{m}=\frac{1}{3}\rho^{m}+\frac{\Pi}{3}.
\end{align}
The electromagnetic lagrangian can be extended up-to second order
terms of the field as \cite{bandyopadhyay2011study}
\begin{equation}\label{a11}
L^{ME}=-\frac{1}{4\mu_{0}}F+{\chi}F^2+{\lambda}F^{*2},
\end{equation}
where $F=F_{\mu\nu}F^{\mu\nu}$, $F^{*}=F^{*}_{\mu\nu}F^{\mu\nu}$ and
$F^{*}$ is the dual of Maxwell electromagnetic field tensor while
$\chi$ and $\lambda$ are the arbitrary constants. The respective
stress-energy tensor is obtained by the extension of stress-energy
tensor of electromagnetic field due to the existence of non-linear
terms in least action, as
\begin{equation}\label{a12}
T^{B}_{\mu\nu}=-4\frac{{\partial}L}{{\partial}F}F_{\mu\eta}F^{\eta}_{\nu}+\bigg(\frac{{\partial}L}{{\partial}F^{*}}F^{*}-L\bigg)g_{\mu\nu}.
\end{equation}
It can be easily verified that $F=(B^2-E^2)$ where $E$ is the
electric field and $B$ is the magnetic field. Now we consider plasma
having homogeneous electric field which give rise to electric
currents in the presence of charged particles. The current rapidly
decays over the passage of time which results in the dominance of
magnetic field over electric field, i.e., $F=2B^2$ and
$<E^2>\approx0$. The energy density $\rho^{B}$ and the pressure
$P^{B}$ for the stress-energy tensor $T^{B}_{\mu\nu}$ can be written
as
\begin{align}\nonumber
\rho^{B}=\frac{B^2}{2\mu_{0}}(1-8\mu_{0}{\chi}B^2)=\frac{B^2}{2\mu_{0}}-4{\chi}B^4,
\\\label{a13}P^{B}=\frac{B^2}{6\mu_{0}}(1-40\mu_{0}{\chi}B^2)=\frac{B^2}{6\mu_{0}}-\frac{40}{6}{\chi}B^4.
\end{align}
The density of the magnetic field should be positive which is
guaranteed by satisfying the condition
$B<\frac{1}{2\sqrt{2\chi\mu_{0}}}$.

The total value of density and pressure can be determined by the
following equations
\begin{align}\label{a14}
&\rho^{total}=\rho^{m}+\rho^{B}=\rho^{m}+\frac{B^2}{2\mu_{0}}-4{\chi}B^4,
\\\label{a15}
&P^{total}=P^{m}+P^{B}=P^{m}+\frac{B^2}{6\mu_{0}}-\frac{40}{6}{\chi}B^4.
\end{align}
Scalar-tensor field variations in the principle action yield the
magnetized CBD field equations as
\begin{equation}\label{a16}
G_{\mu\nu}=T^{(eff)}_{\mu\nu}=\frac{1}{\phi}\bigg(f(\phi)(T^{m}_{\mu\nu}+T^{B}_{\mu\nu})+T^{\phi}_{\mu\nu}\bigg),
\end{equation}
\begin{equation}\label{a17}
\Box\phi=\frac{T}{2w_{CBD}+3}\bigg[f(\phi)-\phi~\frac{\frac{f(\phi)}{\phi}}{2}\bigg]-\frac{2U(\phi)-{\phi}\frac{U(\phi)}{\phi}}{2w_{CBD}+3},
\end{equation}
where ${T}$ gives the trace of stress-energy tensor of mass
distribution as well as for magnetic field, i.e., $T^{m}_{\mu\nu}$,
$T^{B}_{\mu\nu}$ and $\Box$ is $d^{,}$ Alembertian operator and
\begin{equation}\label{a18}
T^{\phi}_{{\mu}{\nu}}=\bigg[\phi_{{,\mu}{;\nu}}-g_{{\mu}{\nu}}\Box\phi\bigg]
-\frac{U(\phi)}{2}g_{{\mu}{\nu}}+\frac{w_{CBD}}{\phi}\bigg[\phi_{,\mu}\phi_{,\nu}-{\frac{1}{2}}g_{{\mu}{\nu}}\phi_{,\omega}\phi^{,\omega}\bigg].
\end{equation}
Here the above expression represents the stress part of scalar field
$\phi$. Equation \eqref{a17} is an evolution equation of scalar
field $\phi$ representing wave equation known as Klein-Gordan
equation. The line element for a spherically symmetric metric
bounded by a hyper-surface $\Sigma^{e}$ is as follows
\begin{equation}\label{a19}
ds^{2}=-A^{2}(t,r)dt^{2}+B^{2}(t,r)dr^{2}+R^{2}(t,r)(d\theta^{2}+\sin^{2}{\theta}d\phi^{2}).
\end{equation}
The magnetized CBD field equations are as follows
\begin{align}\nonumber
&T^{(eff)}_{00}={\frac{1}{\phi}}\bigg(f(\phi)(T^{m}_{00}+T^{B}_{00})+T^{\phi}_{00}\bigg)=\bigg(2\times\frac{\dot{B}}{B}+\frac{\dot{R}}{R}\bigg)\frac{\dot{R}}{R}
-\left(\frac{A}{B}\right)^{2}
\bigg[\bigg(\frac{\acute{R}}{R}\bigg)^{2}+
\\\label{a20}&\times\frac{R''}{R}-2\times \frac{B'}{B} \frac{R'}{R}-
\bigg(\frac{B}{R}\bigg)^{2}\bigg],
\end{align}
where
\begin{align}\nonumber
&{\frac{1}{\phi}}\bigg(f(\phi)(T^{m}_{00}+T^{B}_{00})+T^{\phi}_{00}\bigg)=
\frac{1}{\phi}\bigg[(\rho^{m}+\frac{B^2}{2\mu_{0}}-4{\chi}B^4)
A^{2}-\dot{\phi}\bigg(\frac{\dot{A}}{A}+\frac{\dot{B}}{B}+2\frac{\dot{R}}{R}\bigg)
\\\nonumber&+\phi'\bigg(\frac{A'}{AB}^{2}-\frac{B'}{B}^{3}+\frac{2R'}{RB^{2}}\bigg)\bigg]
+\frac{w_{CBD}}{2\phi^{2}}\bigg(\frac{A^{2}}{\phi'^{2}} +
\dot{\phi}^{2}\bigg)+ \frac{U(\phi)A^{2}}{2\phi} + 2\ddot{\phi} +
\frac{A^{2}\phi''}{B^{2}\phi}, \\\label{a21}
&T^{(eff)}_{01}={\frac{1}{\phi}}\bigg(f(\phi)(T^{m}_{01}+T^{B}_{01})+T^{\phi}_{01}\bigg)=-2\bigg(-\frac{\dot{R}A'}{RA}
-\frac{\dot{BR'}}{BR} + \frac{\dot{R}'}{R}\bigg),
\end{align}
where
\begin{align}\nonumber
&{\frac{1}{\phi}}\bigg(f(\phi)(T^{m}_{01}+T^{B}_{01})+T^{\phi}_{01}\bigg)
=\frac{1}{\phi}\bigg( \dot{\phi}' - qABf(\phi)+
w_{CBD}\bigg(\frac{\dot{\phi}\phi'}{\phi^{2}}\bigg)\bigg),
\\\nonumber
&T^{(eff)}_{11}={\frac{1}{\phi}}\bigg(f(\phi)(T^{m}_{11}+T^{B}_{11})+T^{\phi}_{11}\bigg)=-\bigg(\frac{B}{R}\bigg)^{2}
-\bigg(\frac{B}{A}\bigg)^{2}\bigg[2\frac{\ddot{R}}{R}-\frac{\dot{R}}{R}\bigg(2\frac{\dot{A}}{A}-\frac{\dot{R}}{R}\bigg)\bigg]
+ \\\label{a22}&\bigg(2\frac{A'}{A}+\frac{R'}{R}\bigg)\frac{R'}{R},
\end{align}
where
\begin{align}\nonumber
&{\frac{1}{\phi}}\bigg(f(\phi)(T^{m}_{11}+T^{B}_{11})+T^{\phi}_{11}\bigg)=\frac{1}{\phi}\bigg[f(\phi)(\frac{P^{m}_{r}}{3}+\frac{B^2}{6\mu_{0}}-\frac{40}{6}{\chi}B^4)B^{2}
+\dot{\phi}\bigg(-\frac{\dot{A}}{A^{3}}+\frac{\dot{B}}{A^{2}B}\bigg)+\\\nonumber&\phi'\bigg(-\frac{A'}{AB^{2}}-\frac{2R'}{RB^{2}}+\frac{B'}{B^{3}}\bigg)\bigg]
+\frac{w_{CBD}}{\phi^{2}}\bigg(\phi'^{2}+\frac{B^{2}\dot{\phi}^{2}}{A^{2}}\bigg)-\frac{\ddot{\phi}B^{2}}{{\phi}A^{2}}-\frac{U(\phi)B^{2}}{2\phi},
\\\nonumber &T^{(eff)}_{22}=T^{(eff)}_{33}=
{\frac{1}{\phi}}\bigg(f(\phi)(T^{m}_{22}+T^{B}_{22})+T^{\phi}_{22}\bigg)\\\nonumber&=
{\frac{1}{\phi}}\bigg(f(\phi)(T^{m}_{33}+T^{B}_{22})+T^{\phi}_{33}\bigg)=-\bigg(\frac{R}{A}\bigg)^{2}\bigg[\frac{\ddot{B}}{B}+\frac{\dot{B}\dot{R}}{BR}-\frac{\dot{A}}{A}\bigg(\frac{\dot{B}}{B}+\frac{\dot{R}}{R}\bigg)+\frac{\ddot{R}}{R}\bigg]
\\\label{a23}&+\bigg(\frac{R}{B}\bigg)^{2}\bigg[\frac{A''}{A}-\frac{A'B'}{AB}+\frac{R'}{R}\bigg(\frac{A'}{A}-\frac{B'}{B}\bigg)+\frac{R''}{R}\bigg],
\end{align}
where
\begin{align}\nonumber
&{\frac{1}{\phi}}\bigg(f(\phi)(T^{m}_{22}+T^{B}_{33})+T^{\phi}_{22}\bigg)=
{\frac{1}{\phi}}\bigg(f(\phi)(T^{m}_{33}+T^{B}_{33})+T^{\phi}_{33}\bigg)
\\\nonumber&=\frac{1}{\phi}\bigg[f(\phi)(\frac{2P^{m}_{\bot}}{3}+\frac{B^2}{6\mu_{0}}-\frac{40}{6}{\chi}B^4)R^{2}-\dot{\phi}\bigg(-\frac{\dot{A}}{A}
+\frac{2\dot{R}}{RA^{2}}+\frac{\dot{B}}{A^{2}B}\bigg)\bigg]
\\\nonumber&-\frac{\phi'}{\phi}\bigg(-\frac{A'}{AB^{2}}-\frac{2R'}{R}B^{2}+\frac{B'}{B^{3}}\bigg)\bigg]
+\frac{w_{CBD}}{2\phi^{2}}\bigg(\frac{R^{2}\dot{\phi}^{2}}{A^{2}}-\frac{R^{2}\phi'^{2}}{B^{2}}\bigg)-\frac{\ddot{\phi}R^{2}}{{\phi}A^{2}}-\frac{\phi''R^{2}}{{\phi}B^{2}}
-\frac{U(\phi)R^{2}}{2\phi}.
\end{align}
The wave equation by using Eq. \eqref{a17} is found as
\begin{align}\nonumber
&\frac{\ddot{\phi}}{A^{2}}+\frac{\phi^{''}}{B^{2}}+\phi^{'}\bigg[-\frac{A^{'}}{AB^{2}}+\frac{B^{'}}{B^{3}}-\frac{2R^{'}}{RB^{2}}\bigg]
+\dot{\phi}\bigg[-\frac{\dot{A}}{A^{3}}+\frac{\dot{B}}{AB}+\frac{2\dot{R}}{RA}\bigg]\\&\label{a24}=\frac{\bigg((\frac{P^{m}_{r}}{3}+P^{B}_{r})+2(\frac{2P^{m}_{\bot}}{3}+P^{B}_{\bot})-(\rho^{m}+\rho^{B})\bigg)}{2w_{CBD}+3}\bigg[f(\phi)-
\frac{{\phi}\frac{df(\phi)}{\phi}}{2}\bigg]-\frac{2U(\phi)-{\phi}U(\phi)_{,\phi}}{2w_{CBD}+3}.
\end{align}
The matter configuration can be completely described by three
kinematical quantities namely as 4-acceleration $(a_{\mu})$,
expansion scalar $(\Theta)$ and shear tensor
$(\sigma_{\alpha\zeta})$ . The 4-acceleration is given as
\begin{equation}\label{a25}
a_{\mu}=v_{\mu;\nu}v^{\nu}.
\end{equation}
The above equation for our system gives
\begin{equation}\label{a26}
a_{1}=\frac{A'}{A}, a^{2}=a^{\mu}a_{\mu}=\frac{{A'}^2}{A^2B^2},
\end{equation}
with $a^{\mu}=aX^{\mu}$. The shear tensor is determined as
\begin{equation}\label{a27}
\sigma_{\mu\nu}=a_{(\mu}v_{\nu)}+v_{(\mu;\nu)}-\frac{1}{A}h_{\mu\nu}{\Theta}.
\end{equation}
The non-vanishing components of shear tensor are as follows
\begin{equation}\label{a28}
\sigma_{11}={\frac{2}{3}}{\sigma}B^{2},\quad
\sigma_{22}=\frac{\sigma_{33}}{sin^{2}\phi}=-R^{2}(\frac{1}{3}{\sigma}),
\end{equation}
where
\begin{equation}\nonumber
\sigma^{\mu\nu}\sigma_{\mu\nu}=\frac{2}{3}\sigma^{2},
\end{equation}
and
\begin{equation}\label{a29}
\sigma=\frac{\dot{B}}{AB}-\frac{\dot{R}}{AR}.
\end{equation}
At last, the expansion scalar takes the form
\begin{equation}\label{a30}
\Theta=v^{\mu}_{;\mu}=\frac{\dot{B}}{AB}+\frac{2\dot{R}}{AR}.
\end{equation}

\subsection {The Misner-Sharp mass function and magnetized CBD theory}

The mass function  given by Misner and Sharp
\cite{cahill1970spherical} in case of magnetized CBD theory is
considered as
\begin{equation}\label{a31}
m(t,r)=\frac{R}{2}\bigg[1-\bigg(\frac{R'}{B}\bigg)^{2}+\bigg(\frac{\dot{R}}{A}\bigg)^{2}\bigg].
\end{equation}
The proper derivative $D_{R}$ is proposed as

\begin{equation}\label{a32}
D_{R}=\frac{1}{R'}(\frac{\partial}{\partial{r}}).
\end{equation}
Similarly, proper time derivative $D_{T}$ is given as
\begin{equation}\label{a33}
D_{T}=\frac{1}{A}(\frac{\partial}{\partial{t}}).
\end{equation}
The velocity of collapsing fluid after using Eq. \eqref{a33} is as
follows
\begin{equation}\label{a34}
V=D_{T}R=\frac{1}{A}(\frac{{\partial}R}{{\partial}t}).
\end{equation}
Putting Eq. \eqref{a34} in Eq. \eqref{a31}, it results in
\begin{equation}\label{a35}
E=\bigg[V^2+1-\frac{2m}{R}\bigg]^{\frac{1}{2}}=\frac{R'}{B}.
\end{equation}
With Eq. \eqref{a31} we can get
\begin{equation}\label{a36}
D_{R}m=\frac{1}{2\phi}R^{2}\bigg[(\rho^{m}+\frac{B^2}{2\mu_{0}}-4{\chi}B^4){f(\phi)}+\frac{V}{E}\bigg(-\frac{T^{\phi}_{01}}{AB}+qf(\phi)\bigg)+\frac{T^{\phi}_{00}}{A^{2}}\bigg].
\end{equation}
By integrating Eq. \eqref{a36}, we have
\begin{equation}\label{a37}
m=\int^{r}_{0}\frac{1}{2\phi}R^{2}\bigg[(\rho^{m}+\frac{B^2}{2\mu_{0}}-4{\chi}B^4)+\frac{V}{E}\bigg(q-\frac{T^{\phi}_{01}}{BA}\bigg)\bigg]R'dr.
\end{equation}
In spherically symmetric models, electric part $(E_{\mu\nu})$ of
conformal tensor is
\begin{equation}\label{a38}
E_{\mu\nu}=\epsilon\bigg[\chi_{\mu}\chi_{\nu}-\frac{h_{\mu\nu}}{3}\bigg],
\end{equation}
where
\begin{align}\label{a39}
\varepsilon&=-\frac{1}{2R^{2}}+\frac{1}{2A^{2}}\bigg[\frac{\ddot{R}}{R}-\bigg(\frac{\dot{A}}{A}
+\frac{\dot{R}}{R}\bigg)\bigg(\frac{\dot{R}}{R}-\frac{\dot{B}}{B}\bigg)
-\frac{\ddot{B}}{B}\bigg]\\\nonumber&+\frac{1}{2B^{2}}\bigg[\frac{A''}{A}
+\bigg(\frac{R'}{R}-\frac{A'}{A}\bigg)\bigg(\frac{B'}{B}+\frac{R'}{R}\bigg)-\frac{R''}{R}\bigg].
\end{align}
From magnetized CBD field equations, mass function and conformal
tensor, the relationship between these three quantities can be
determined as
\begin{equation}\label{a40}
\frac{3m(t,r)}{R^{3}}=\frac{1}{\phi}\bigg[f(\phi)\frac{[(\rho^{m}+\frac{B^2}{2\mu_{0}}-4{\chi}B^4)-\Pi]}{2}-\varepsilon+\frac{1}{2}\bigg(3\Box\phi+
\frac{w_{CBD}\phi_{,\gamma}\phi^{,\gamma}}{\phi}\bigg)\bigg].
\end{equation}
{The above equation shows the effect of magnetic field in the mass
function along with the contribution of $f(\phi)$ function and
conformal tensor. The equation \eqref{a40} shows the effects of magnetic field in the
mass function along with  the contribution of $f(\phi)$ function and
conformal tensor with the embodiment of the basic features of the
matter, such as, energy density irregularity and pressure
anisotropy.}

\subsection{ Orthogonal Splitting of the Riemann tensor in Magnetized CBD theory}

Herrera \emph{et al.} \cite{herrera2004spherically} performed
orthogonal splitting of the curvature tensor by considering it for
radiating as well as non-radiating galactic structures in GR. From
the orthogonal splitting technique we can also determine scalar
functions known as structure scalars given by $Y_{T}, Y_{TF}, X_{T},
X_{TF}$ . Thus by considering Bel technique
\cite{bel1961inductions}, tensors have the form
\begin{align}\label{a41}
&Y_{\mu\nu}=R_{\mu\alpha\nu\gamma}u^{\alpha}u^{\beta},\quad
X_{\mu\nu}=^*R^*_{\mu\alpha\nu\beta}u^{\alpha}u^{\beta}.
\end{align}
The other form of the above expressions can be written as
\begin{equation}\label{a42}
Y_{\mu\nu}={\frac{1}{3}}Y^{a}_{a}h_{\mu\nu}+Y_{<\mu\nu>},
\end{equation}
where
\begin{equation}\label{a43}
Y_{T}=Y^{a}_{a},
\end{equation}
and
\begin{equation}\label{a44}
Y_{<\mu\nu>}=Y^{m}_{TF}f(\phi)\bigg(\chi_{\mu}\chi_{\nu}-\frac{1}{3}h_{\mu\nu}\bigg)
+Y^{\phi}_{TF}\bigg(\frac{g_{\mu\nu}}{4}-\frac{h_{\mu\nu}}{3}\bigg).
\end{equation}
The trace-free part of Eq. \eqref{a42} is

\begin{equation}\label{a45}
Y_{TF}=Y^{m}_{TF}f(\phi)+Y^{\phi}_{TF},
\end{equation}

\begin{equation}\label{a46}
Y^{m}_{TF}=\frac{1}{\phi}\bigg(\varepsilon-\frac{\Pi}{2}\bigg),
\end{equation}
and
\begin{equation}\label{a47}
Y^{\phi}_{TF}=\frac{1}{\phi}\bigg[\frac{\Box\phi}{2}-\phi_{,\zeta;\beta}u^{\zeta}u^{\beta}
-\frac{w_{CBD}}{\phi}\phi_{,\mu}\phi_{,\beta}u^{\mu}u^{\beta}+\frac{w_{CBD}}{2\phi}\phi_{,\alpha}\phi^{\beta}+\frac{U(\phi)}{2}\bigg].
\end{equation}
In the same manner
\begin{equation}\label{a48}
X_{\mu\nu}={\frac{1}{3}}X^{a}_{a}h_{\mu\nu}+X_{<\mu\nu>},
\end{equation}
where
\begin{equation}\label{a49}
X_{T}=X^{a}_{a},
\end{equation}
\begin{equation}\label{a50}
X_{<\mu\nu>}=X^{m}_{TF}f(\phi)\bigg(\chi_{\mu}\chi_{\nu}-\frac{1}{3}h_{\mu\nu}\bigg)
 +X^{\phi}_{TF}\bigg(\frac{g_{\mu\nu}}{4}-\frac{h_{\mu\nu}}{3}\bigg),
\end{equation}
with
\begin{equation}\label{a51}
X^{m}_{TF}=\frac{1}{\phi}\bigg(-\varepsilon-\frac{\Pi}{2}\bigg),
\end{equation}
and
\begin{equation}\label{a52}
X^{\phi}_{TF}=-\frac{1}{\phi}\bigg[\frac{\Box\phi}{2}-\phi_{,\zeta;\beta}u^{\zeta}u^{\beta}
+\frac{w_{CBD}}{2\phi}\phi_{,\alpha}\phi^{\beta}-\frac{w_{CBD}}{\phi}\phi_{,\mu}\phi_{,\beta}u^{\mu}u^{\beta}\bigg].
\end{equation}
The respective values of Eq. \eqref{a43} and Eq. \eqref{a49} are as
follows
\begin{align}\nonumber
&Y_{T}=Y^{m}_{T}f(\phi)+Y^{\phi}_{T}=
\frac{f(\phi)}{2\phi}\bigg((\rho^{m}+\frac{B^2}{2\mu_{0}}-4{\chi}B^4)-2\Pi+3(P^{m}+\frac{B^2}{6\mu_{0}}
-\frac{40}{6}{\chi}B^4)\bigg)\\\label{a53}&-\frac{1}{\phi}\bigg[\frac{\Box\phi}{2}-\phi_{,\alpha;\beta}u^{\alpha}u^{\beta}
-\frac{w_{CBD}}{\phi}\phi_{,\mu}\phi_{,\beta}u^{\mu}u^{\beta}-\frac{U(\phi)}{2}\bigg].
\end{align}

\begin{align}\nonumber
&X_{T}=X^{m}_{T}f(\phi)+X^{\phi}_{T}=\frac{f(\phi)(\rho^{m}+\frac{B^2}{2\mu_{0}}-4{\chi}B^4)}{\phi}+\frac{1}{\phi}\bigg(\Box\phi+\phi_{,\zeta;\beta}u^{\zeta}u^{\beta}+\frac{U(\phi)}{2}\bigg)
\\\label{a54}&+\frac{w_{CBD}}{2\phi}\bigg(\phi^{\alpha}\phi_{\alpha}+\phi_{,\zeta;\beta}u^{\zeta}u^{\beta}\bigg).
\end{align}
The relationship between mass function and structure scalars is
elaborated as
\begin{equation}\label{a55}
\frac{3m}{R^{3}}=\frac{1}{\phi}\bigg[3X^{m}_{TF}f(\phi)+\frac{3X^{m}_{T}}{2}-\bigg(Y^{\phi}_{T}+2Y^{\phi}_{TF}\bigg)\bigg].
\end{equation}
In literature, $Y_{TF}$ has been found to be a complexity factor
\cite{herrera2018definition, herrera2019complexity1,
herrera2019complexity, herrera2020stability,maurya2022role}. The
complexity factor combining with scalar functions along with the
kinematical quantities reduces to the form
\begin{equation}\label{a56}
\frac{a'}{B}-\frac{\dot{\sigma}}{A}-\frac{2\Theta\sigma}{3}-\frac{aR'}{RB}-\frac{\sigma^{2}}{3}+a^{2}
=\frac{1}{\phi}\bigg[-f(\phi)Y^{m}_{TF}+\frac{3Y^{\phi}_{TF}}{8}+\frac{Y^{\phi}_{T}}{4}-X^{\phi}_{TF}+\frac{X^{\phi}_{T}}{2}\bigg].
\end{equation}
Our galactic models would become non-complex if it satisfy the
minimal complexity condition, i.e, $Y_{TF}=0$. {With the help of $Y_{TF}$, it is possible to determine how anisotropy
and inhomogeneity affect the system. The expression is described as
follows}
\begin{align}\nonumber
&Y_{TF}=-\Pi
f(\phi)+\frac{1}{2R^3}\int^{r}_{0}\bigg[f(\phi)\bigg(\rho'+(\frac{B^2}{2\mu_{0}}-4{\chi}B^4)'\bigg)-\frac{3q}{RBA}\frac{V}{E}\bigg]R'R^3dr
+\bigg[\bigg(\frac{2\dot{A}}{A}+\frac{3\dot{R}}{R}\bigg)\frac{\dot{\phi}}{A^2}\bigg]
\\&\label{a55a}-\frac{\phi'}{B^2}\frac{3R'}{R}
-\frac{T^{\phi}_{22}}{R^2}+\frac{T^{\phi}_{11}}{B^2}.
\end{align}
{The aforementioned equation reflects the impact of physical
characteristics like irregular energy density ($\rho'$) and pressure
anisotropy ($\Pi$) caused by $Y_{TF}$ and the dark source terms of
the modified gravity in the presence of magnetic field effects.
Therefore, $Y_{TF}$ conveys the maximum information require to
identify the system's complexity. The above equation shows that the
complexity of the system increases in the influence of non-linear
electromagnetic field. As a result, $Y_{TF}$ effectively determines
the complexity level of the astrophysical object.}

\section{Matching conditions for Magnetized CBD gravity}

This section is comprised with matching conditions in order to avoid
the existence of shells on the exterior surface delimiting the
models in magnetized CBD theory \cite{bonnor1981junction,
brustein2022classical}. If the model describing the fluid
distribution have a cavity surrounded the center then Darmois
conditions must be imposed. Since some of the models permit the
emergence of cavity, in this scenario, we consider the matching on
both the metrics i.e., inner as well as outer. Outside $\Sigma^{e}$,
we have Vadiya metric, described by
\begin{equation}\label{a57}
ds^{2}=-\bigg[1-\frac{2M(v)}{r}\bigg]-2drdv+r^{2}(d\theta^{2}+\sin^{2}{\theta}d\phi^{2}),
\end{equation}
where $M(v)$ and $v$ describes the total mass and retardation time,
respectively. Within magnetized CBD gravity, to match the
non-adiabatic spherical surface with the Vaidya space-time metric,
implies the continuity of the first and second differential forms.
In case of non-existence of shells on the boundary surface
$r=r_{\Sigma^({e})}$=constant, we obtain
\begin{equation}\label{a58}
m(t,r)\overset{\Sigma^{(e)}}= M(v),
\end{equation}
and
\begin{equation}\label{a59}
q\overset{\Sigma^{(e)}}=\frac{L}{4{\pi}r}\overset{\Sigma^{(e)}}=P^{m}_{r}+P^{EM}_{r},
 \end{equation}
where $L$ describes the total luminance of the spherically symmetric
configuration and $\overset{\Sigma^{(e)}}=$ represents that
expressions are evaluated on the boundary $\Sigma^{(e)}$. The value
of $L$ is given by
\begin{equation}\label{a60}
L=L_{\infty}\bigg(1+2\frac{dr}{dv}-\frac{2m}{r}\bigg)^{-1},
\end{equation}
with
\begin{equation}\label{a61}
L_{\infty}=\frac{dM(v)}{dv},
\end{equation}
where $L_{\infty}$ is the value of total luminosity measured by an
observer at infinity. When a void is formed, Minkowski metric is
matched with the hyper-surface by delimiting the empty void
$\Sigma^{(i)}$. Thus the junction (matching) conditions become
\begin{equation}\label{a62}
m(t,r)\overset{\Sigma^{(e)}}=0,
\end{equation}
\begin{equation}\label{a63}
q\overset{\Sigma^{(e)}}=P^{m}_{r}+P^{EM}_{r}=0.
\end{equation}
For some particular models Darmois conditions cannot be justified by
showing discontinuities in Misner-Sharp mass. For this, existence of
 thin shells must be allowed on hyper-surfaces $\Sigma^{(i)}$ and
 $\Sigma^{(e)}$.

\section{Quasi-homologous $(Q_h)$ condition in Magnetized CBD theory}

For some time dependent systems, complexity factor is not enough to
describe their structures. For this the dynamical system is in need
of another condition which was described by an easier pattern of
evolution, known as $Q_{h}$ condition \cite{herrera2020quasi}. Here
we shall relax this condition by eliminating some potential
scenarios. By $Q_{h}$ condition, the magnetized CBD field equation
\eqref{a21} can be written as
\begin{equation}\label{a64}
D_R\bigg(\frac{V}{R}\bigg)=\frac{1}{{\phi}E}\bigg(f(\phi)q-\frac{T^{\phi}_{01}}{AB}\bigg)+\frac{\sigma}{R},
\end{equation}
which on integration results in
\begin{equation}\label{a65}
V=\check{o}(t)R+\int^{r}_{0}RR'\bigg[\frac{1}{{\phi}E}\bigg(f(\phi)q-\frac{T^{\phi}_{01}}{AB}\bigg)+\frac{\sigma}{R}\bigg]dr,
\end{equation}
where $\check{o}(t)$ is an integration constant or one can write
\begin{equation}\label{a66}
V=\frac{V_{\Sigma^{(e)}}}{R_{\Sigma^{(e)}}}R-\int^{r_{\Sigma^{(e)}}}_{r}RR'\bigg[\frac{1}{{\phi}E}\bigg(f(\phi)q-\frac{T^{\phi}_{01}}{AB}\bigg)+\frac{\sigma}{R}\bigg]dr.
\end{equation}
If the integral vanishes then from  Eq. \eqref{a65}, we obtain
\begin{equation}\label{a67}
V=\check{o}(t)R.
\end{equation}
\begin{equation}\label{a67}
V=\check{o}(t)R.
\end{equation}
Let's take a look at two concentric shells, characterized by $r=r_
{1}=constant$ and $r=r_{2}=constant$, respectively. We have
\begin{equation}\label{b2}
\frac{T_{I}}{T_{II}}=constant.
\end{equation}
The factors listed in Eqs. \eqref{a67} and \eqref{b2} interpret the
system's homologous evolution. {The fundamental problem we wish to
underline that the requirements in the Eqs.  \eqref{a67} and \eqref
{b2} are distinct from one another. For two fluid shells, Eq.
\eqref{a67} implies}
\begin{equation}\label{b3}
\frac{T_{I}}{T_{II}}=\frac{G_{II}
\dot{R_{I}}}{G_{I}\dot{R_{I}}}=\frac{R_{I}}{R_{II}}.
\end{equation}
{It entails Eq. \eqref{b2} only if $G = G(t)$, which becomes A =
constant by a basic coordinate transformation. As a nutshell, in the
non-relativistic domain, the condition that the distance is
equivalent to the radial velocity always follows from Eq. \eqref{b2}
but in the relativistic zone, the condition \eqref{a67} entails
\eqref{b2} only if the fluid is geodesic. This is known as $Q_{h}$
constraint.} By using Eq. \eqref{a67}, ${Q_h}$ can be reconsidered as
\begin{equation}\label{a68}
\frac{1}{{\phi}E}\bigg(f(\phi)q-\frac{T^{\phi}_{01}}{AB}\bigg)+\frac{\sigma}{R}=0.
\end{equation}
Thus our stellar models will be confined by $Q_h$ condition and
$Y_{TF}=0$.

\section{Kinematical Conditions and Radial Velocity}

Irrespective of the vanishing $Y_{TF}$  and $Q_h$ conditions, we
have to impose new conditions to examine stellar structures in
magnetized CBD theory \cite{yousaf2021quasi}. For this, a different
definition of velocity not same as $V$ along with extra restrictions
is determined. The new notion of velocity can be taken as extremely
small proper radial distance between two neighboring points per unit
proper time $t$. It can be written as
\begin{equation}\label{a69}
\frac{D_T({\delta}l)}{{\delta}l}=\frac{\Theta}{3}+\frac{2\sigma}{3},
\end{equation}
or
\begin{equation}\label{a70}
\frac{D_T({\delta}l)}{{\delta}l}=\frac{\dot{B}}{BA}.
\end{equation}
Then we can write
\begin{equation}\label{a71}
\sigma=-\frac{D_TR}{R}+\frac{D_T({\delta}l)}{{\delta}l}=-\bigg(\frac{V}{R}\bigg)+\frac{D_T({\delta}l)}{{\delta}l},
\end{equation}
and
\begin{equation}\label{a72}
\Theta=\frac{2D_TR}{R}+\frac{D_T({\delta}l)}{{\delta}l}=2\bigg(\frac{V}{R}\bigg)+\frac{D_T({\delta}l)}{{\delta}l},
\end{equation}
where $\frac{D_T({\delta}l)}{{\delta}l}$ describes the velocity of
the fluid and $V$ is connected with areal radius $R$. By considering
the case when $\frac{D_T({\delta}l)}{{\delta}l}=0$ but $V\neq0$ then
$B=B(r)$. Without loss of generality we can take $B=1$  and $R'=E$
\cite{herrera2008shearing}. From Eq. \eqref{a71} and Eq.
\eqref{a72}, we have
\begin{equation}\label{a73}
\sigma=-\frac{V}{R}=-\frac{\Theta}{2}.
\end{equation}
Equation \eqref{a64} can be stated as
\begin{equation}\label{a74}
\frac{{\sigma}R'}{R}+\sigma'=\frac{T^\phi_{01}}{\phi}-\frac{f(\phi)q}{\phi}.
\end{equation}
On integration of above equation with respect to  $r$, we obtain
\begin{equation}\label{a75}
\sigma=\frac{1}{R}\int^r_0\frac{1}{\phi}\bigg(T^\phi_{01}-f(\phi)q\bigg)Rdr+\frac{\psi(t)}{r},
\end{equation}
where $\psi(t)$ is the integration constant. From Eq. \eqref{a75},
we have
\begin{equation}\label{a76}
V=-\psi(t)+\int^r_0\frac{1}{\phi}\bigg(T^\phi_{01}-f(\phi)q\bigg)Rdr.
\end{equation}
If the sphere has no void then $\psi=0$ then Eq. \eqref{a76} becomes
\begin{equation}\label{a77}
V=\int^r_0\frac{1}{\phi}\bigg(T^\phi_{01}-f(\phi)q\bigg)Rdr.
\end{equation}
When sphere has a cavity then $\psi$ has value different from zero.
When $q>0$ i.e., when direction of flux vector is outward, from Eq.
\eqref{a72} and Eq. \eqref{a77}, we have $\Theta>0$, $V>0$ (all
terms within the integral are positive). When $q<0$, it results in
the dissipation of energy which contracts during the
Kelvin-Helmholtz phase. The above discussion suggests that
$\psi\neq0$. Thus it can be easily seen that the new areal velocity
condition is suitable for describing the evolving fluid distribution
having a void bounding the center. Another kinematical restriction
of fluid distribution is $V=0$ and $D_T({\delta}l)\neq0$. In this
case, infinitesimal proper radial distance changes over time whereas
areal radius remains the same. From Eqs. \eqref{a70} and
\eqref{a71}, $Q_h$ condition \eqref{a68} can be rewritten as
\begin{equation}\label{a78}
\frac{\dot{B}}{AB}=-\frac{1}{{\phi}E}\bigg(-\frac{T^\phi_{01}}{AB}+f(\phi)q\bigg),\quad\quad
\sigma=\Theta.
\end{equation}
The above kinematical condition does not assure the formation of
void covering the center of fluid distribution. In the next section,
we analyze galactic models satisfying the vanishing complexity
condition, i.e., $Y_{TF}=0$.

\section{ Transport equation in CBD Gravity}

In diffusion approximation method, to calculate temperature as well
as its evolution in fluid distribution with respect to magnetized
CBD theory, we require a transport equation. The respective equation
was derived from a known dissipative theory (Mullar-Israel-Stewart
Theory) \cite{muller1967paradoxon}. The equation for heat flux can
be illustrated as
\begin{equation}\label{a79}
{\tau}h^{\mu\nu}q_{\nu;\gamma}\check{V^{\gamma}}+q^\mu=-\kappa(T_,\nu+Ta_\nu)h^{\mu\nu}-\frac{1}{2}\bigg(\frac{\tau\check{V^\gamma}}{{\kappa}T^2}\bigg)_{;\nu}{q^\mu},
\end{equation}
where $T$, $\kappa$ and $\tau$ represent temperature, thermal
conductivity and proper time, respectively. The non-vanishing
independent component of the above equation is written as
\begin{equation}\label{a80}
{\tau}\dot{q}^{(eff)}=-\frac{1}{2}{\tau}q^{(eff)}\bigg(\frac{\tau}{{\kappa}T^2}\bigg)T^2-Aq^{(eff)}-\frac{\kappa}{B}\bigg(TA\bigg)'
-\frac{1}{2}{\tau}q^{(eff)}\Lambda\Theta.
\end{equation}
For our convenience, the new version of Eq. \eqref{a79} neglecting
the last part in it can be of the form
\begin{equation}\label{a81}
{\tau}h^{\mu\nu}q_{\nu;\gamma}\check{V^{\gamma}}+q^\mu=-\kappa(T_{,\nu}+Ta_\nu)h^{\mu\nu},
\end{equation}
where the only non-zero independent component comes out as
\begin{equation}\label{a82}
{\tau}\dot{q}^{(eff)}+Aq^{(eff)}=-\frac{\kappa}{B}\frac{\partial{(TA)}}{\partial{r}}.
\end{equation}

\section{Galactic models in Magnetized CBD theory}

In this section, we shall determine the exact solutions of
dissipative as well as non-dissipative models satisfying the minimal
complexity condition, i.e., $Y_{TF}$=0. The systems will evolve by
$Q_h$ condition whereas further restrictions will be used to specify
the models.

\subsection{Non-dissipative model $(q=0)$}

In this case, model is satisfying the condition, i.e., $Y_{TF}$=0
which shows that the fluid is geodesic. From Eq. \eqref{a68}, if
$q=0$ then by this condition $\sigma=0$ which results in
\begin{equation}\label{a83}
R=B(t,r)r.
\end{equation}
By the above expression, Eq. \eqref{a21} becomes
\begin{equation}\label{a84}
\bigg(\frac{\dot{B}}{AB}\bigg)'+\frac{1}{2\phi}(\frac{T^\phi_{01}}{A})=0.
\end{equation}
Then Eq. \eqref{a67} can be written as
\begin{equation}\label{a85}
V=\frac{\dot{R}}{A}=\frac{\dot{B}r}{A}=\check{o}(t)Br,
\end{equation}
where the above equation satisfies the magnetized CBD field
equations. In this scenario, if we restrict the model, i.e., $B=1$
or $V=0$ \cite{herrera2018definition, herrera2018tilted}, a static
spherical model would be produced. Applying $Y_{TF}=0$ by using Eq.
\eqref{a56} shows that $\sigma=0$, we obtain
\begin{equation}\label{a86}
\frac{a'}{B}+a^{2}-\frac{aR'}{RB}=\frac{1}{\phi}\bigg[\frac{Y^\phi_T}{4}-X^\phi_{TF}+\frac{X^\phi_T}{2}\bigg],
\end{equation}
or, by using Eq.\eqref{a26}, we have
\begin{equation}\label{a87}
A''+\frac{2B'A'}{B}-\frac{A'}{r}=\frac{1}{\phi}\bigg[AB^2\bigg(\frac{Y^\phi_T}{4}-X^\phi_{TF}+\frac{X^\phi_T}{2}\bigg)\bigg],
\end{equation}
which on integration yields as
\begin{equation}\label{a88}
A'=P^\phi\frac{R^2}{r}K(t),
\end{equation}
where
\begin{equation}\nonumber
P^\phi=e^{\int^r_0\frac{1}{\phi}\frac{B^2A}{A'}\bigg(\frac{Y^\phi_T}{4}-X^\phi_{TF}+\frac{X^\phi_T}{2}\bigg)dr},
\end{equation}
here $K(t)$ is an integration function. Now by using Eqs.
\eqref{a83}, \eqref{a85} and \eqref{a88}, Eq. \eqref{a84} can be
rewritten as
\begin{equation}\label{a89}
\frac{\dot{R'}}{R}-\frac{\dot{R}R'}{R^2}=\frac{R^2}{r}\check{}{o}(t)K(t)P^\phi-\frac{T^\phi_{01}}{2\phi}.
\end{equation}
Introducing $g=\frac{R'}{R}$, Eq. \eqref{a89} can be revised as
\begin{equation}\label{a90}
\dot{g'}=\dot{g}\bigg(2g-\frac{1}{r}\bigg)+\frac{T^\phi_{01}}{2\phi}-\frac{1}{2}\bigg(\frac{T^\phi_{01}}{\phi}\bigg)'.
\end{equation}
By considering new variable $h=gr$, above equation reduces to
\begin{equation}\label{a91}
\dot{h'}=\frac{2h\dot{h}}{r}+r\bigg[\frac{T^\phi_{01}}{2\phi}-\frac{1}{2}\bigg(\frac{T^\phi_{01}}{\phi}\bigg)'\bigg],
\end{equation}
or, taking the independent variable $x={\ln}r$, Eq. \eqref{a91}
becomes
\begin{equation}\label{a92}
\frac{d\dot{h}}{dx}=2h\dot{h}+\bigg[\frac{T^\phi_{01}}{2\phi}-\frac{1}{2}\bigg(\frac{T^\phi_{01}}{\phi}\bigg)'\bigg].
\end{equation}
By using Mathematica, above equation can be solved as
\begin{equation}\label{a93}
h=-c_2\tanh(c_3+c_2\ln{r}+c_1t),
\end{equation}
or, by using Eq. \eqref{a93}, we may have
\begin{equation}\label{a94}
R=\frac{\check{R}(t)\varsigma^\phi}{{\cosh}u},
\end{equation}
where $u=c_3+c_2\ln{r}+c_1t$, $\check{R}(t)$ and $\varsigma^\phi$
are integration functions and $c_1, c_2, c_3$ are integration
constants. The value $\varsigma^\phi$ is given as
\begin{equation}\nonumber
\varsigma^\phi=e^{\int^r_0\frac{1}{2r}\bigg[\frac{T^\phi_{01}}{\phi}-\bigg(\frac{T^\phi_{01}}{\phi}\bigg)'\bigg]dr}.
\end{equation}
In order to simplify the model, we take $\check{o}(t)$=constant, the
field equations are
\begin{equation}\label{a95}
\frac{f(\phi)(\rho^m+\frac{B^2}{2\mu_{0}}-4{\chi}B^4)}{\phi}+\frac{T^\phi_{00}}{{\phi}A^2}=3{\check{o}}^2+\frac{1}{{\check{R}^2(t)}\varsigma^{\phi^2}}\bigg[(1-c_2^2)\cosh^2+3c_2^2\bigg],
\end{equation}
\begin{align}\nonumber
&\frac{f(\phi)(\frac{P^{m}_{r}}{3}+\frac{B^2}{6\mu_{0}}-\frac{40}{6}{\chi}B^4)}{\phi}+\frac{T^\phi_{11}}{{\phi}B^2}=-3{\check{o}}^2
+\frac{1}{{\check{R}^2(t)}\varsigma^{\phi^2}D}\bigg[\dot{\check{R}}(t)\bigg(-c_2^2+(c_2^2-1)\cosh^2\bigg)
\\\label{a96}&+\check{R}(t)c_1{{\tanh}u}\bigg(3c_2^2+(1-c_2^2\cosh^2u)\bigg)\bigg]+3\frac{\dot{\phi}}{\phi}c_2^2{\sinh^2}u\check{R}(t)
-\frac{\dot{\phi}}{\phi}{\cosh^2}u\check{R}(t),
\\\label{a97}
&\frac{f(\phi)(\frac{2P^{m}_\bot}{3}+\frac{B^2}{6\mu_{0}}-\frac{40}{6}{\chi}B^4)}{\phi}+\frac{T^\phi_{22}}{{\phi}R^2}=
-3{\breve{o}}^2+\frac{c_2^2}{{\breve{R}^2(t)}\varsigma^{\phi^2}D}
\bigg[3{\breve{R}(t)}{\tanh}u-\dot{\breve{R}}(t)+\frac{\dot{\phi}}{\phi}\breve{R}(t)\bigg],
\end{align}
where
\begin{equation}\label{a98}
D\equiv\dot{\check{R}}(t)-\check{R}(t){\tanh}uc_1+\frac{\dot{\varsigma^\phi}}{\varsigma^\phi}\check{R}(t).
\end{equation}
For large range of  values of parameters, the above explanation is
singular-free satisfying the usual energy as well as boundary
conditions. This model clarifies that with $Q_h$ condition, it
becomes an ease to justify $Y_{TF}=0$.

\subsection{ Dissipative cases $(V\neq0, D_T({\delta}l)=0)$}

We now consider the models satisfying $D_T({\delta}l)$=0 which gives
$B=1$. In these models, the center of the system is endowed with a
void(cavity). In this scenario, field equations become
\begin{align}\label{a99}
&\frac{1}{\phi}\bigg[f(\phi)(\rho^{m}+\frac{B^2}{2\mu_{0}}-4{\chi}B^4)+\frac{T^\phi_{00}}{A^2}\bigg]=
\frac{1}{A^2}\frac{\dot{R}^2}{R^2}-\frac{R'^2}{R^2}-\frac{2R''}{R}+\frac{1}{R^2},
\\\label{a100}&\frac{1}{2\phi}\bigg[qf(\phi)-\frac{T^\phi_{01}}{A}\bigg]=\frac{1}{A}\bigg(\frac{\dot{R}'}{R}-\frac{A'}{A}\frac{\dot{R}}{R}\bigg)=-\sigma\frac{R'}{R},\\\nonumber
&\frac{1}{\phi}\bigg[f(\phi)(\frac{P^{m}_{r}}{3}+\frac{B^2}{6\mu_{0}}-\frac{40}{6}{\chi}B^4)+\frac{T^\phi_{11}}{B^2}\bigg]=
-\frac{1}{A^2}\bigg(2\frac{\ddot{R}}{R}-2\frac{\dot{A}}{A}\frac{\dot{R}}{R}+\frac{\dot{R^2}}{R^2}\bigg)
\\\label{a101}&+\frac{R'^2}{R^2}+2\frac{A'}{A}\frac{R'}{R}-\frac{1}{R^2},
\\\label{a102}&\frac{1}{\phi}\bigg[f(\phi)(\frac{2P^{m}_\bot}{3}+\frac{B^2}{6\mu_{0}}-\frac{40}{6}{\chi}B^4)+\frac{T^\phi_{22}}{R^2}\bigg]=
-\frac{1}{A^2}\bigg(\frac{\ddot{R}}{R}-\frac{\dot{A}}{A}\frac{\dot{R}}{R}\bigg)
+\frac{A''}{A}+\frac{A'}{A}\frac{R'}{R}+\frac{R''}{R}.
\end{align}
The kinematical variables for above equations are
\begin{equation}\label{a103}
\sigma=-\frac{1}{RA}(\frac{\partial{R}}{\partial{t}}), \quad
\Theta=\frac{2}{RA}(\frac{\partial{R}}{\partial{t}}).
\end{equation}
Now using $Q_h$ condition, we have
\begin{align}\label{a104}
&V=\check{o}(t)R=\frac{\dot{R}}{A}\quad\Rightarrow\quad
\check{o}(t)=\frac{\dot{R}}{RA}\quad\Rightarrow\quad
\sigma=\check{o}(t),\\\label{a105}& \Theta-\sigma=3\check{o}(t).
\end{align}
The condition $Y_{TF}=0$ results in
\begin{equation}\label{a106}
Y_{TF}=\frac{A''}{A}+\frac{1}{A^2}\bigg(\frac{\ddot{R}}{R}-\frac{\dot{A}}{A}\frac{\dot{R}}{R}\bigg)-\frac{R'A'}{RA}
=\frac{1}{\phi}\bigg[\frac{Y^\phi_T}{4}+\frac{X^\phi_T}{2}-X^\phi_{TF}\bigg].
\end{equation}
\begin{equation}\label{a107}
\sigma^2-\frac{\dot{\sigma}}{A}-\frac{R'A'}{RA}+\frac{A''}{A}-\frac{\xi^\phi}{\phi}=0,
\end{equation}
where
\begin{equation}\nonumber
\frac{\xi^\phi}{\phi}=\frac{1}{\phi}\bigg[\frac{Y^\phi_T}{4}+\frac{X^\phi_T}{2}-X^\phi_{TF}\bigg].
\end{equation}
By the $Y_{TF}=0$ and $Q_h$ condition, previous expressions can be
read as
\begin{equation}\label{a108}
A''-\frac{R'A'}{R}+A\sigma^2=\dot{\sigma}+\frac{A}{\phi}{\xi^\phi},
\end{equation}
and
\begin{equation}\label{a109}
\frac{\dot{R}}{R}=-(\sigma{A}),
\end{equation}
respectively. By introducing intermediate variables, one can write
\begin{equation}\label{a110}
A=X+\frac{\dot{\sigma}}{\sigma^2}\quad and \quad
R=Y\frac{\partial{X}}{\partial{r}}.
\end{equation}
Now Eqs. \eqref{a108} and \eqref{a109} can become
\begin{align}\label{a111}
&\frac{\dot{X}'}{X'}+\frac{\dot{Y}}{Y}=-\frac{\dot{\sigma}}{\sigma}-X{\sigma},
\\\label{a112}
&-\frac{X'Y'}{XY}+\sigma^2=\frac{A}{X\phi}{\xi^\phi}.
\end{align}
By imposing further restrictions, stellar models can be analyzed
justifying the above conditions.

\subsubsection{Case with $X=\check{X}(r)\check{T}(t)$}

Let us consider a separable function $X$, i.e.,
\begin{equation}\label{a113}
X=\check{X}(r)\check{T}(t).
\end{equation}
By putting Eq. \eqref{a113} into Eq. \eqref{a112} and taking time
derivative, we have
\begin{equation}\label{a114}
-\frac{\check{X}'}{\check{X}}\bigg(\frac{\dot{Y}}{Y}\bigg)'+2\sigma\dot{\sigma}=\bigg(\frac{A}{\check{X}\check{T}\phi}{\xi^\phi}\bigg)_{,0}.
\end{equation}
By again inserting Eq. \eqref{a113} into Eq. \eqref{a111}, we obtain
\begin{equation}\label{a115}
\bigg(\frac{\dot{Y}}{Y}\bigg)'=-\sigma\check{X}'\check{T}.
\end{equation}
By merging the Eqs. \eqref{a114} and \eqref{a115}, we get
\begin{equation}\label{a116}
\frac{\check{X}'^2}{\check{X}}=\frac{1}{\check{T}}\bigg[-2\dot{\sigma}+\bigg(\frac{A}{\check{X}\check{T}\phi}{\xi^\phi}\bigg)_{,0}\bigg]=\eta^2,
\end{equation}
where $\eta$ is a constant value. Integration of  Eq. \eqref{a116}
gives
\begin{equation}\label{a117}
\check{X}=\frac{\bigg({\eta}r+\acute{c}\bigg)^2}{4}\quad and \quad
\check{T(t)}=\frac{-2\dot{\sigma}+\bigg(\frac{A}{\check{X}\check{T}\phi}{\xi^\phi}\bigg)_{,0}}{\eta^2},
\end{equation}
where $\acute{c}$ is an integration constant. Thus metric functions
in this case are
\begin{align}\label{a118}
&A=\frac{\dot{\sigma}}{2\eta^2\sigma^2}\bigg[2\eta^2-\sigma^2\bigg({\eta}r+\acute{c}\bigg)^2\bigg]+
\frac{({\eta}r+\acute{c})^2}{4\sigma\eta}\bigg(\frac{A}{\check{X}\check{T}\phi}{\xi^\phi}\bigg)_{,0},
\\\label{a119}
&R=\check{R}(t)\frac{\eta}{2}\bigg({\eta}r+\acute{c}\bigg)e^{\frac{\sigma^2r}{4\eta}({\eta}r+2\acute{c})+Z^\phi_1},
\end{align}
where
\begin{equation}\nonumber
Z^\phi_1=\int\frac{r^2}{4}\bigg(\frac{A}{\check{X}\check{T}\phi}{\xi^\phi}\bigg)'dr-\frac{r^2}{4}\bigg(\frac{A}{\check{X}\check{T}\phi}{\xi^\phi}\bigg)
-\frac{\acute{c}}{2\eta}\int\bigg(\frac{A}{\check{X}\check{T}\phi}{\xi^\phi}\bigg)dr,
\end{equation}
and $\check{R}(t)$ is an arbitrary time function. By using Eqs.
\eqref{a99}-\eqref{a102} in above equations, the physical variables
are
\begin{align}\nonumber
&\frac{1}{\phi}\bigg[f(\phi)(\rho^{m}+\frac{B^2}{2\mu_{0}}-4{\chi}B^4)+T^\phi_{00}\bigg]=
-3\sigma^2-\frac{3\sigma^4}{4\eta^2}({\eta}r+\acute{c})^2
-\frac{\eta^2}{({\eta}r+\acute{c})^2}\\\label{a120}&+\frac{e^{-\frac{\sigma^2r}{4\eta}({\eta}r+2\acute{c})
-Z^{\phi}_1}}{\check{R^2}(t)\frac{\eta^2}{4}({\eta}r+\acute{c})^2}
+\varphi^\phi_1, \\\label{a121}
&\frac{1}{\phi}\bigg[f(\phi)q+T^\phi_{01}\bigg]=-\sigma\bigg[\frac{2\eta^2+2\eta({\eta}r
+\acute{c})Z^{\phi'}_1+\sigma^2({\eta}r+\acute{c})^2}{\eta({\eta}r+\acute{c})}\bigg],\\\nonumber
&\frac{1}{\phi}\bigg[f(\phi)(\frac{P^{m}_{r}}{3}+\frac{B^2}{6\mu_{0}}-\frac{40}{6}{\chi}B^4)+T^\phi_{11}\bigg]
=-\frac{4\sigma^2\eta^2}{\bigg[2\eta^2-\sigma^2({\eta}r+\acute{c})^2+\frac{\sigma^2}{\dot{\sigma}\eta^2}
({\eta}r+\acute{c})^2\dot{(\frac{A\xi^\phi}{{\phi}X})}\bigg]}
\\\label{a122}&+\frac{\eta^2}{({\eta}r+\acute{c})^2}+\frac{\sigma^4}{4\eta^2}({\eta}r+\acute{c})^2
-\frac{-e^{\frac{\sigma^2r}{4\eta}({\eta}r+2\acute{c})+Z^{\phi}_1}}{\breve{R^2}(t)\frac{\eta^2}{4}({\eta}r+\acute{c})^2}
+\varphi^\phi_2, \\\nonumber
&\frac{1}{\phi}\bigg[f(\phi)(\frac{2P^{m}_\bot}{3}+\frac{B^2}{6\mu_{0}}-\frac{40}{6}{\chi}B^4)+T^\phi_{22}\bigg]=
-\frac{\sigma^2\bigg[2\eta^2+\sigma^2({\eta}r+\acute{c})^2\bigg]}{\bigg[2\eta^2-\sigma^2({\eta}r+\acute{c})^2
+\frac{\sigma^2}{\dot{\sigma}\eta^2}({\eta}r+\acute{c})^2\bigg(\dot{\frac{A\xi^\phi}{{\phi}X}}\bigg)\bigg]}
\\\label{a123}&+\frac{\sigma^4}{4\eta^2}({\eta}r+\acute{c})^2+\frac{\sigma^2}{2}+\varphi^\phi_3.
\end{align}
The values of $\varphi^\phi_1$, $\varphi^\phi_2$, $\varphi^\phi_3$
are given in Appendix A. Now for the matching condition on inner and
outer surfaces, we have $\sigma=0$ on the boundary surface with
constructs a non-dissipative system due to dependence on $t$ only.
In view of the junction (Darmois) condition, we have
\begin{equation}\label{a124}
\frac{\sigma(2+\sigma^2H^2)}{2H}-\frac{4\sigma^2}{2-\sigma^2H^2}+\frac{1}{H^2}+\frac{\sigma^4H^2}{4}-\frac{\sigma^2}{\eta^2H^2}e^{-\frac{\sigma^2H^2}{2}}=0,
\end{equation}
where the value of $H$ is
\begin{equation}\nonumber
H=\frac{{\eta}r\sigma^{(e)}+\acute{c}}{\eta}.
\end{equation}
The above statement shows that solutions only exist for those values
of $t$ which remains fixed i.e., values of $\sigma$ are constant
depending on ($H$, $\eta$). At last, we can evaluate the value of
temperature by using Eqs. \eqref{a118} and \eqref{a121} which
results in
\begin{align}\nonumber
&T(t,r)=-\frac{2\sigma^2\eta^2}{\kappa\bigg(\dot{\sigma}\bigg[2\eta^2-\sigma^2\bigg({\eta}r+\acute{c}\bigg)^2\bigg]+
\frac{({\eta}r+\acute{c})^2}{4\sigma\eta}\bigg(\frac{A}{\check{X}\check{T}\phi}{\xi^\phi}\bigg)_{,0}\bigg)}\\\nonumber&
\varrho\int\bigg[\dot{\bigg(\frac{T^\phi_{01}}{Af(\phi)}\bigg)}-\dot{\bigg(\frac{\sigma}{f(\phi)}
\bigg[\frac{2\eta^2+\sigma^2({\eta}r+\acute{c})+2\eta({\eta}r+\acute{c})Z^\phi_1}{2\eta}\bigg]\bigg)}\bigg]
dr\\\nonumber&+\int\bigg[\frac{T^\phi_{01}}{f(\phi)}-\frac{\sigma}{2{\eta}f(\phi)}\bigg(2\eta^2+\sigma^2({\eta}r+\acute{c})^2
+2\eta({\eta}r+\acute{c}){Z^\phi_1}'\bigg)
\\\label{a125}&\frac{\dot{\sigma}}{2\eta^2\sigma^2}\bigg[2\eta^2-\sigma^2\bigg({\eta}r+\acute{c}\bigg)^2\bigg]+
\frac{({\eta}r+\acute{c})^2}{4\sigma\eta}\bigg(\frac{A}{\check{X}\check{T}\phi}{\xi^\phi}\bigg)_{,0}\bigg]dr.
\end{align}

\subsubsection{Case with $A=A(r)$}

 In this case we assume that$A=A(r)$.
Taking $r$-derivative of Eq. \eqref{a109}, we obtain

\begin{equation}\label{a126}
\bigg(\frac{\dot{R}}{R}\bigg)'=-A'{\sigma},
\end{equation}
whereas $t$-derivative of Eq. \eqref{a108} results in
\begin{equation}\label{a127}
-A'\bigg(\frac{\dot{R}}{R}\bigg)'+2A\sigma\dot{\sigma}=\ddot{\sigma}+A\dot{\bigg(\frac{\xi^\phi}{\phi}\bigg)}.
\end{equation}
By combining the Eqs. \eqref{a126} and \eqref{a127} we have
\begin{equation}\label{a128}
\sigma(A')^2+2A\sigma\dot{\sigma}=\ddot{\sigma}+A\dot{\bigg(\frac{\xi^\phi}{\phi}\bigg)},
\end{equation}
where the solution of the above equation can be expressed as
\begin{equation}\label{a129}
A=\frac{1}{4}\bigg[\sqrt{2}L(r)+c\bigg]^2,
\end{equation}
where
\begin{equation}\nonumber
L(r)=\int{\sqrt{(\xi^\phi)+\frac{1}{\sigma}\dot{(\xi^\phi)}}}dr,
\end{equation}
and
\begin{equation}\label{a130}
\sigma=-\bigg((\xi^\phi)+\sigma_0\bigg)+\sigma_1,
\end{equation}
where $\sigma_0$, $\sigma_1$ reads as constants. By considering
above expressions, $R$ produces as
\begin{equation}\label{a131}
R=\bar{R}(r)e^{-\frac{1}{4}(-\frac{\sigma-o}{t^2}+\sigma_1t-y(t))(\sqrt{2}L(r)+c)^2},
\end{equation}
where $\bar{R}(r)$ is an arbitrary function. For this particular
case, we assume $\bar{R}(r)=0$. By interchanging Eq. \eqref{a130}
and Eq. \eqref{a131} in eqs. \eqref{a99}-\eqref{a102}, we have field
equations as
\begin{align}\nonumber
&\frac{1}{\phi}\bigg[f(\phi)(\rho^{m}+\frac{B^2}{2\mu_{0}}-4{\chi}B^4)+\frac{T^\phi_{00}}{A^2}\bigg]=\bigg(-{\sigma_0}t+\sigma_1-\dot{y(t)}\bigg)^2
-\sqrt{2}(-\sigma_o\frac{t^2}{2}+\sigma_1t\\\nonumber&-y(t))\bigg[\sqrt{2}(L'(r))^2+L''(r)(\sqrt{2}L(r)+c)-\frac{\sqrt{2}}{2}(\sqrt{2}L(r)+c)L(r)
(-\sigma_o\frac{t^2}{2}+\sigma_1t-y(t))\bigg]\\\label{a132}&+\frac{1}{2}(L'(r))^2(-\sigma_o\frac{t^2}{2}+\sigma_1t-y(t))(\sqrt{2}L(r)+c)
+\frac{1}{\bar{R}(r)^2}e^{\frac{1}{2}(\sqrt{2}L(r)+c)^2(-\sigma_o\frac{t^2}{2}+\sigma_1t-y(t))},
\\\label{a133}
&\frac{1}{2\phi}\bigg[f(\phi)q+\frac{T^\phi_{01}}{A}\bigg]=(-\sigma_o\frac{t^2}{2}+\sigma_1t-y(t))\bigg[\frac{1}{\sqrt{2}}{L(r)}'(\sqrt{2}L(r)+c)\bigg],
\\\nonumber&\frac{1}{\phi}\bigg[f(\phi)(\frac{P^{m}_{r}}{3}+\frac{B^2}{6\mu_{0}}-\frac{40}{6}{\chi}B^4)+\frac{T^\phi_{11}}{B^2}\bigg]=+\frac{8\bar{R}(r)
(-\sigma_o-\ddot{g(t)})}{(\sqrt{2}L(r)+c)^2}
-2\bar{R}(r)(-{\sigma_0}t+\sigma_1-\dot{y(t)})^2\\\nonumber&-(-{\sigma_0}t+\sigma_1-\dot{y(t)})^2-4(L'(r))^2(\sigma_o\frac{t^2}{2}+\sigma_1t-y(t))
+\frac{1}{2}(L'(r))^2(\sigma_o\frac{t^2}{2}+\sigma_1t-y(t))^2\\\label{a134}&(\sqrt{2}L(r)+c)^2
-\frac{1}{\bar{R}(r)^2}{e}^{\frac{1}{2}(\sqrt{2}L(r)+c)^2(-\sigma_o\frac{t^2}{2}+\sigma_1t-y(t))},
\\\nonumber&\frac{1}{\phi}\bigg[f(\phi)(\frac{2P^{m}_\bot}{3}+\frac{B^2}{6\mu_{0}}-\frac{40}{6}{\chi}B^4)+\frac{T^\phi_{22}}{R^2}\bigg]=\frac{4\bar{R}(r)(-\sigma_o-
\ddot{y(t)})}{(\sqrt{2}L(r)+c)^2}
-\bar{R}(r)(-{\sigma_0}t+\sigma_1-\dot{y(t)})^2\\\nonumber&-\frac{\sqrt{2}}{2}(-\sigma_o\frac{t^2}{2}+\sigma_1t-y(t))\bigg[\sqrt{2}(L'(r))^2+P''(r)(\sqrt{2}L(r)+c)
\\\nonumber&-\frac{\sqrt{2}}{2}(\sqrt{2}L(r)+c)L(r)(-\sigma_o\frac{t^2}{2}+\sigma_1t-y(t))\bigg]-2(L'(r))^2(\sigma_o\frac{t^2}{2}+\sigma_1t-y(t))
\\\label{a135}&\frac{2\sqrt{2}L''(r)}{(\sqrt{2}L(r)+c)^2}+\frac{4(L(r)')^2}{(\sqrt{2}L(r)+c)}.
\end{align}
Now we check the matching conditions on inner as well as on boundary
surfaces. For radial pressure Eq. \eqref{a134}, regularity
conditions shows that $(\sqrt{2}L(r)+c)\neq0$, then the matching
conditions implies that $\sigma=0$, developing a non-dissipation
solution.

The above discussion can be completed by evaluating the temperature
of this model by using Eqs. \eqref{a129} and \eqref{a133} as
\begin{align}\nonumber
&T(t,r)=\frac{-4}{\kappa(\sqrt{2}L(r)+c)^2}\int\bigg[\varrho\bigg[\dot{\frac{\phi}{f(\phi)}\bigg(-\frac{\sigma_ot^2}{2}+\sigma_1t
-y(t)\bigg)\bigg(\frac{L(r)'(\sqrt{2}L(r)+c)}{2}\bigg)}\\\nonumber&+\dot{\frac{T^\phi_{01}}{f(\phi)A}}\bigg]
\frac{T^\phi_{01}}{f(\phi)}+\frac{\phi}{4f(\phi)}\bigg(\frac{-\sigma_ot^2}{2}+\sigma_1t-y(t)\bigg)\bigg(\frac{\sqrt{2}L(r)'(\sqrt{2L(r)+c})}{2}\bigg)
\\\label{a136}&\bigg(\sqrt{2}L(r)+c\bigg)^2\bigg]dr.
\end{align}

\subsubsection{Case when $\dot{\sigma}=0$}
As $\dot{\sigma}=0$ which shows $\sigma$=constant. Now by presenting
the intermediate variable $Y$ we have
\begin{equation}\label{a137}
R=YA',
\end{equation}
where Eqs. \eqref{a108}, \eqref{a109} can be of the form
\begin{equation}\label{a138}
(\frac{Y'}{Y})(\frac{A'}{A})=\sigma^2-\frac{\xi^\phi}{\phi},
\end{equation}
and
\begin{equation}\label{a139}
\frac{\frac{dY}{dt}}{Y}+\frac{\dot{A}'}{A}=-A\sigma.
\end{equation}
 Now by taking derivative with respect to $t$ of Eq. \eqref{a138} and derivative with
 respect to $r$ of Eq. \eqref{a139}, we get
 \begin{equation}\label{a140}
 \bigg(\frac{\frac{dY}{dt}}{Y}\bigg)'=-\bigg(\frac{\frac{dA}{dt}}{A}\bigg)'\bigg(\frac{A}{A'}\bigg)^2\bigg[\sigma^2
 -\dot{\bigg(\frac{1}{\phi}{\xi^\phi}\bigg)}\bigg]
 -\dot{\bigg(\frac{1}{\phi}{\xi^\phi}\bigg)},
 \end{equation}
and
\begin{equation}\label{a141}
\bigg(\frac{\frac{dY}{dt}}{Y}\bigg)'=-{\sigma}A'-\bigg(\frac{(\frac{dA}{dt})'}{A'}\bigg).
\end{equation}
combining the above expressions gives us
\begin{equation}\label{a142}
{\sigma}A'^3+\dot{A}''A'-\dot{A}'A''+\bigg(\sigma^2-\dot{(\frac{\phi}{\xi^\phi})}\bigg)(A'\dot{A}-\dot{A}'A)+A'^2\dot{\bigg(\frac{\phi}{\xi^\phi}\bigg)}=0.
\end{equation}
A solution of the above expression is of the form
\begin{equation}\label{a143}
A={\psi}r-t\frac{\psi^2}{\sigma}+\omega_0,
\end{equation}
where
\begin{equation}\nonumber
\psi=\omega+\frac{\xi^\phi}{\phi}.
\end{equation}
Now by examining above equations, $R$ produces as
\begin{equation}\label{a144}
R=\breve{R_o}~{\psi}~e^{\sigma^2\frac{r^2}{2}-\sigma{\psi}tr
+\sigma^2r\frac{\omega_o}{\psi}+\frac{t^2}{2}\psi^2-\sigma_ot+Y^\phi},
\end{equation}
where $\breve{R_o}$ is just an arbitrary function. By plugging Eq.
\eqref{a143} and Eq. \eqref{a144} in Eqs. \eqref{a99}-\eqref{a102},
the field equations become
\begin{align}\label{a145}
&\frac{1}{\phi}\bigg[f(\phi)(\rho^{m}+\frac{B^2}{2\mu_{0}}-4{\chi}B^4)+T^\phi_{00}\bigg]=-\sigma^2-3\bigg[\sigma^2r-t\sigma\psi
+\sigma^2\frac{\omega_o}{\psi}+{Y^\phi}'\bigg]+\lambda_{1}^{\phi},
\\\label{a146}
&\frac{1}{\phi}\bigg[f(\phi)q+T^\phi_{01}\bigg]=-\sigma^3\bigg[r-t\frac{\psi}{\sigma}+t\frac{\omega_o}{\psi}
+\frac{{Y^\phi}'}{\sigma}\bigg],\\\label{a147}
&\frac{1}{\phi}\bigg[f(\phi)(\frac{P^{m}_{r}}{3}+\frac{B^2}{6\mu_{0}}-\frac{40}{6}{\chi}B^4)+T^\phi_{11}\bigg]=-\sigma^2-2\frac{\sigma^2}{{\psi}r-
t\frac{\psi^2}{\sigma}+\omega_0}+\lambda_{2}^{\phi},
\\\label{a148}&
\frac{1}{\phi}\bigg[f(\phi)(\frac{2P^{m}_\bot}{3}+\frac{B^2}{6\mu_{0}}-\frac{40}{6}{\chi}B^4)+
\frac{T^\phi_{22}}{R^2}\bigg]=\sigma^2+\frac{\dot{Y^\phi}}{\sigma^2\bigg({\psi}r-
t\frac{{\psi}^2}{\sigma}+\omega_0\bigg)^3}
\psi^2+\lambda_{3}^{\phi}.
\end{align}
In this case, junction conditions are not justified for both
surfaces, i.e., for inner $\Sigma^{(i)}$ surface as well as for
boundary surface $\Sigma^{(e)}$. The temperature in this case can be
evaluated as
\begin{align}\nonumber
&T(t,r)=-\frac{1}{\kappa({\psi}r-t\frac{{\psi}^2}{\sigma}+\omega_0)}
\int\bigg[\bigg(\dot{\frac{T^\phi_{01}}{f(\phi)A}}\bigg)-\dot{\bigg[\frac{2\phi}{f(\phi)}\sigma^3(r-t\frac{{\psi}}{\sigma}
+\frac{\omega_o}{{\psi}}+\frac{{Y^\phi}'}{\sigma})\bigg]}\\\label{a149}&+\frac{T^\phi_{01}}{f(\phi)}
-\bigg[\frac{2\phi}{f(\phi)}\sigma^3(r-t\frac{{\psi}}{\sigma}
+\frac{\omega_o}{{\psi}}+\frac{{Y^\phi}'}{\sigma})\bigg]
{\psi}r-t\frac{{\psi}^2}{\sigma}+\omega_0\bigg)\bigg]dr+\breve{T_o}(t).
\end{align}
Lastly, we investigate the case for $V=0$.

\subsection{dissipative cases $V=0$, $D_T({\delta}l)\neq0$}

Considering $V=0$, it follows that
\begin{equation}\label{a150}
V=\frac{1}{A}\frac{\partial{R}}{\partial{t}}=0\Longrightarrow
R=R(r).
\end{equation}
Then the kinematical quantities become
\begin{equation}\label{a151}
\Theta=\sigma=\frac{1}{BA}\frac{\partial{B}}{\partial{t}}.
\end{equation}
By combining above expressions with the equations Eqs. \eqref{a56},
\eqref{a68}, we have
\begin{equation}\label{a152}
\frac{1}{B^2}\bigg(\frac{A''}{A}-\frac{A'B'}{AB}-\frac{A'R'}{AR}\bigg)
=\frac{1}{A^2}\bigg(\frac{\ddot{B}}{B}-\frac{\dot{B}\dot{A}}{BA}\bigg)+\frac{\xi^\phi}{\phi}.
\end{equation}
A model can be created satisfying Eq. \eqref{a152}, by assuming
\begin{equation}\label{a153}
\frac{A''}{A}-\frac{A'B'}{AB}-\frac{A'R'}{AR}=0,
\end{equation}
\begin{equation}\label{a154}
\ddot{B}-\frac{\dot{B}\dot{A}}{A}=\frac{A^2B\xi^\phi}{\phi},
\end{equation}
which on integration yields as
\begin{equation}\label{a155}
A'=BRK(t),
\end{equation}
\begin{equation}\label{a156}
\dot{B}=Ak(r)\varrho^\phi,
\end{equation}
where
\begin{equation}\nonumber
\varrho^\phi=e^{\int\frac{AB\xi^\phi}{\phi}}dt.
\end{equation}
Here $k(r)$ and $K(t)$ are integration functions. Now, by taking
derivative with respect to $t$ of Eq. \eqref{a156} and derivative
with respect to $r$ Eq. \eqref{a155}, we obtain
\begin{align}\label{a157}
&\frac{\dot{A}'}{A}=\frac{A'\dot{K}}{AK}+RKk,
\\\label{a158}
&\frac{\dot{B}'}{B}=BRKk\xi^\phi+\frac{\dot{B}}{B}\frac{(k\varrho^\phi)'}{k\varrho^\phi}.
\end{align}
By adjoining the above equations, we have
\begin{equation}\label{a159}
\frac{\dot{A}'}{A}-\frac{\dot{B}'}{B}=\frac{A'\dot{K}}{AK}-\frac{\dot{B}}{B}\frac{(k\varrho^\phi)'}{k\varrho^\phi},
\end{equation}
whose solution is
\begin{equation}\label{a160}
A=\frac{CBK}{k\varrho^\phi},
\end{equation}
where $C$ is a constant of integration. Substituting Eq.
\eqref{a160} into Eq. \eqref{a156}, we have
\begin{equation}\label{a161}
\frac{\dot{B}}{B}=CK\Longrightarrow B=B_1(t)B_2(r).
\end{equation}
By the re-parametrization of time function, $A=A(r)$. By taking
$t$-derivative of Eq. \eqref{a160}, we obtain
\begin{equation}\label{a162}
\frac{1}{K}\frac{\partial{K}}{\partial{t}}=-\frac{1}{B}\frac{\partial{B}}{\partial{t}}.
\end{equation}
By using Eqs. \eqref{a156}, \eqref{a160}, we get
\begin{equation}\label{a163}
\frac{\dot{B}}{B}=KC.
\end{equation}
By plugging the above equation into Eq. \eqref{a162} and integrating
with respect to time gives
\begin{equation}\label{a164}
K=\frac{1}{\gamma+Ct},
\end{equation}
where $\gamma$ is an integration constant. Again injecting Eq.
\eqref{a160} into Eq. \eqref{a155}, we have
\begin{equation}\label{a165} \frac{A'}{A}=\frac{kR}{C}.
\end{equation}
The expression for $B$ can be obtained by inserting Eq. \eqref{a160}
into Eq. \eqref{a164} as
\begin{equation}\label{a166}
B=\frac{Ak\varrho^\phi(\gamma+Ct)}{C},
\end{equation}
while from Eqs. \eqref{a151}, \eqref{a160} and \eqref{a163}, the
value of  shear stress tensor can be written as
\begin{equation}\label{a167}
\sigma=\frac{k\varrho^\phi}{B}.
\end{equation}
By considering the above equations, physical variables are
\begin{align}\label{a168}
&\frac{f(\phi)(\rho^{m}+\frac{B^2}{2\mu_{0}}-4{\chi}B^4)}{\phi}=-\frac{\sigma^2}{k^2\varrho{^\phi}^2}\bigg[2\frac{R''}{R}+\frac{R'^2}{R^2}-
\frac{2R'}{R}\bigg(\frac{Rk\varrho^\phi}{C}+\frac{k'}{k}
+\frac{{\varrho^\phi}'}{\varrho^\phi}\bigg)\bigg]+\frac{1}{R^2}-\frac{T^\phi_{00}}{{\phi}A^2},
\\\label{a169}
&\frac{f(\phi)q}{\phi}=\frac{T^\phi_{01}}{{\phi}AB}-2\frac{\sigma^2}{k\varrho^\phi}\frac{R'}{R},
\\\label{a170}
&\frac{f(\phi)(\frac{P^{m}_{r}}{3}+\frac{B^2}{6\mu_{0}}-\frac{40}{6}{\chi}B^4)}{\phi}=\frac{\sigma^2}{k^2\varrho{^\phi}^2}\bigg[2\frac{Rk\varrho^\phi}{C}\frac{R'}{R}
+2\frac{R'^2}{R^2}\bigg]-\frac{1}{R^2}-\frac{T^\phi_{11}}{{\phi}B^2},
\\\label{a171}
&\frac{f(\phi)(\frac{2P^{m}_\bot}{3}+\frac{B^2}{6\mu_{0}}-\frac{40}{6}{\chi}B^4)}{\phi}=\frac{\sigma^2}{k^2\varrho{^\phi}^2}\bigg[\frac{R'k\varrho^\phi}{C}+\frac{R''}{R}-
\bigg(\frac{k'}{k}+\frac{{\varrho^\phi}'}{\varrho^\phi}\bigg)\frac{R'}{R}\bigg]-\frac{T^\phi_{22}}{\phi{R^2}}.
\end{align}
Now to simply the model, we suppose
\begin{equation}\label{a172}
A=R^nb \quad \Rightarrow k=\frac{nCR'}{R^2\varrho^\phi},
\end{equation}
where $n,C$ and $b$ are constants. Thus the equations become
\begin{align}\label{a173}
&\frac{f(\phi)(\rho^{m}+\frac{B^2}{2\mu_{0}}-4{\chi}B^4)}{\phi}=\frac{\sigma^2R^2}{n^2C^2}(2n-5)+\frac{1}{R^2}-\frac{T^\phi_{00}}{{\phi}A^2},
\\\label{a174}
&\frac{f(\phi)q}{\phi}=\frac{T^\phi_{01}}{{\phi}BA}-2\frac{\sigma^2R}{Cn},
\\\label{a175}
&\frac{f(\phi)(\frac{P^{m}_{r}}{3}+\frac{B^2}{6\mu_{0}}-\frac{40}{6}{\chi}B^4)}{\phi}=\frac{\sigma^2R^2}{n^2C^2}(2n+1)-\frac{1}{R^2}-\frac{T^\phi_{11}}{{\phi}B^2},
\\\label{a176}
&\frac{f(\phi)(\frac{2P^{m}_\bot}{3}+\frac{B^2}{6\mu_{0}}-\frac{40}{6}{\chi}B^4)}{\phi}=\frac{\sigma^2R^2}{n^2C^2}(n+2)-\frac{T^\phi_{22}}{\phi{R^2}}.
\end{align}
In this model, there is no need of the Minkowskian cavity covering
the center. If this situation arises then $\sigma=0$ which results
in a non-dissipative stellar model. We can conclude the model by
describing the temperature which can be obtained by using eqs.
\eqref{a82}, \eqref{a172}, \eqref{a174} as
\begin{align}\nonumber
&T(t,r)=-\frac{1}{\kappa{R^nb}}\int\bigg[\tau\dot{\bigg[\frac{T^\phi_{01}}{f(\phi)b^2R^2n\varrho^\phi(Ct+\gamma)}-
\frac{2\phi}{f(\phi)}\frac{1}{nb^2R^2n}{(Ct+\gamma)^2}bR^nk\varrho^\phi(Ct+\gamma)\bigg]}
\\\label{a177}&+\bigg[\frac{T^\phi_{01}}{f(\phi)bR^n{\varrho^\phi}k(Ct+\gamma)}-\frac{2\phi}{f(\phi)}\frac{R}{nbR^n(Ct+\gamma)^2}\bigg]dr+T_0(t).
\end{align}

\section{Conclusions}
This paper is devoted to understand the affects of CBD theory on the
self gravitating systems in the presence of non-linear
electrodynamics. In the realm of astrophysics, self-gravitating
fluid distribution has such intriguing qualities due to which
researches are eager to investigate its physical characteristics
such as pressure, temperature, density, etc. To carry out this
analysis, we have taken into account the anisotropic fluid
configuration associated with matter contents in the presence of
non-linear magnetic effects. To understand the basic fluid
characteristics, we examined certain kinematical variables including
expansion scalar, four-acceleration, shear tensor etc. The modified
field equations are then evaluated while keeping magnetic CBD theory
in mind. In order to match the inner and boundary surfaces smoothly,
matching conditions in the presence of magnetic field must be
achieved. Furthermore, the Misner-Sharp mass function is derived to
determine the amount of matter contained in a spherically symmetric
fluid configuration. An interaction between Misner-Sharp mass,
conformal tensor and metric functions have been established because
of their enormous importance in the analysis of stellar
configurations.

We have also studied the outcomes of the vanishing complexity
condition $(Y_{TF})$ combined with $Q_h$ condition with the magnetic
effects under the influence of the chameleonic field. We have also
considered additional kinematical constraints to produce various
specific models. One of these conditions has been found which is
suitable for depicting the evolution of fluid configuration with a
void covering the center. Some models satisfy matching (Darmois)
condition preventing the presence of shells on delimiting
hyper-surfaces. Other models emerge when Darmois conditions are
weakened and Israel conditions are applied across shells.
Dissipative and non-dissipative galactic models are considered
satisfying the $Q_h$ condition with minimal complexity factor in the
effect of non-linear electromagnetic field. In a non-linear
electromagnetic field, the density of the anisotropic fluid with the
magnetic effects remained the same but the pressure stresses changed
with an additional term in it. Some of the important findings
demonstrating the outcomes of non-linear electrodynamics in CBD
gravity are as follows

$\bullet$ We have evaluated the mass function for homogeneous as
well as non-homogeneous density in terms of magnetic field. The
structure scalar ($Y_{TF}$) has also participated in determining the
changes emerging in $m(t,r)$.

$\bullet$ The conformal tensor and intrinsic curvature, which are
responsible for the formation of tidal forces, are also assessed.
These quantities are eventually related with fluid distribution and
metric variables with the inclusion of dark source terms emerging
from the magnetic CBD theory.

$\bullet$ Under the action of additional curvature terms due to
magnetic field, $Y_{TF}$ covers the anisotropic pressure and
inhomogeneity of energy density in a specified way.

$\bullet$ The effects of magnetic field on the possible existence of
theoretical $Q_h$ spherical models in CBD theory is the framework of
this paper.

$\bullet$ We used the averaging procedure of electromagnetic field
to calculate the CBD gravity owing to the existence of non-linear
terms. For this purpose, the lagrangian in this case is extended
up-to second-order.

$\bullet$ The extension of the energy-momentum tensor due to
electromagnetic field in the presence of non-linear terms is an
appropriate way to determine the effect of energy density and
pressure anisotropy in our paper.

$\bullet$ For the non-dissipative model (non-static) Sec. VII (A),
$Y_{TF}$ gives the dissipative fluxes as well as inhomogeneity of
energy density and local anisotropy in the addition of modified
terms in the magnetic CBD theory. We further apply the condition of
minimal value of complexity factor. By the $Q_h$ and $Y_{TF}$=0
conditions, we get exact analytical solutions of the CBD theory in
the presence of electromagnetic field representing collapsing fluid
systems, whereas in GR, we get analytical solutions describing
collapsing fluid spheres. For non- dissipative anisotropic fluid
configuration $(q=0)$, fluid is geodesic as well as shear-free.

$\bullet$ In case of dissipative collapsing anisotropic fluid
configuration Sec. VII (C) when V=0, center is empty. Scalar
potential function and $f(\phi)$ along with dynamical variables are
used in the evolution of such distribution in the presence of
magnetic field whereas in GR, governing equations involve dynamical
quantities (mass function, conformal tensor, etc.). The $Y_{TF}=0$
condition in dissipative models give exact analytical solutions
involving scalar field in magnetic CBD theory. For this case, fluid
is geodesic but shearing.

{The homologous criteria seems overly restrictive because it results
in a separate model in the non-dissipative case. In fact, the
homologous criterion infers $Y_{TF}$ in this case. This results in
the simplest arrangement, which is only develops homogeneously and
reaches $Y_{TF}=0$. The significant factor we want to make is that
we don't bother about any certain celestial model. However, we want
to clarify that once the homologous restriction is relaxed and the
$Q_{h}$ constraint is assumed, numerous cosmological solutions that
satisfy the requirement $Y_{TF} = 0$ are possible. The simulation of
cosmic voids is one potential use for the presented data. Voids are
the cosmological formations that are neither empty nor spherical,
but are rather thought to be the under-density areas examined in the
large-scale matter distribution in our universe
\cite{gottlober2003structure,zeldovich1982giant}.}

Finally, from the aforementioned results we deduced that in magnetic
CBD gravity, all dynamical alternations in an evolving system are
determined by the scalar field in the presence of magnetic effect
instead of the structural properties (conformal tensor, shear
stress, etc.). The density inhomogeneity, shear-free
characteristics, etc. are all present in the derived models of
modified gravity for a constant scalar field. Otherwise, scalar
distribution (irregular) retains dynamical features such as pressure
anisotropy and its effects, inhomogeneous nature of density,
dissipation, etc., results in the diversion from GR. The Einstein
gravity can be achieved when scalar field is constant
$(\phi=\phi_{0})$, $w\rightarrow\infty$ and $T\neq0$.

{Voids typically described as having spherical vacuum holes
surrounded by a fluid. The hypothesis of a spherically symmetric
space-time outside the hole is feasible for holes with a dimension
of $20M pc$ or less. The impact of voids in astrophysics study is
shown by the observation that the real cosmos appears to have a
structure resembling a sponge that is affected by voids. This figure
is supported by data showing that a cavity having an appropriate
size of f $30h^{-1}M pc$, $H_{0} = 100hkms^{-1}M pc^{-1}$, makes up
approximately half of the universe's volume. Consequently, a range
of cavities, from mini-voids \cite{tikhonov2009sizes} to super-voids
\cite{rudnick2007extragalactic} may be found. The purpose of this
manuscript  is not to build specific models of any known cavity, but
rather to provide crucial equations and raise awareness of the
notion that the only valid criterion for such models is evolutional
constraint.}

\section*{Data Availability Statement}

{All data generated or analyzed during this study are included in this published article.}

\section*{Acknowledgments}

The work of KB was partially supported by the JSPS KAKENHI Grant
Number 21K03547.

\vspace{0.5cm}


\vspace{0.3cm}

\renewcommand{\theequation}{A\arabic{equation}}
\setcounter{equation}{0}
\section*{Appendix A}

The expressions $\varphi_{1}^{\phi}$ , $\varphi_{2}^{\phi}$ and
$\varphi_{3}^{\phi}$ emerging in equations \eqref{a120},
\eqref{a122} and \eqref{a123} are given as
\begin{align}\nonumber
&\varphi_{1}^{\phi}=\frac{{Z^\phi_1}'^2}{4\eta^2}[r\eta^2-1]-\frac{5\sigma^2}{4\sigma^2}({\eta}r+\acute{c}){Z^\phi_1}'
-\frac{{Z^\phi_1}'}{({\eta}r+\acute{c})}-2\frac{\sigma^2}{\eta}({\eta}r+\acute{c})^2{Z^\phi_1}'-{Z^\phi_1}''.
\\\nonumber
&\varphi_{2}^{\phi}=\frac{\frac{2\sigma^4}{\eta^2}\bigg[\frac{({\eta}r+\acute{c})}{\dot{\sigma}}\bigg(\dot{\frac{A}{{\phi}X}}\xi^\phi\bigg)+\frac{2\eta^3}{\sigma^2}({\eta}r+\acute{c}){Z^\phi_1}'\bigg]}
{\bigg[2\sigma^2-\sigma^2({\eta}r+\acute{c})^2+\frac{\sigma^2}{\dot{\sigma}\eta^2}\bigg(\dot{\frac{A}{{\phi}X}\xi^\phi}\bigg)\bigg]}
\\\nonumber&-\frac{\eta^3\bigg[2\eta^2+\sigma^2({\eta}r+\acute{c})^2+2\eta({\eta}r+\acute{c}){Z^\phi_1}'\bigg[\frac{({\eta}r+\acute{c})^2}{4\eta^2}\dot{(\frac{A}{{\phi}X}\xi^\phi)}'\bigg]\bigg]}
{\sigma^2\dot{\sigma}({\eta}r+\acute{c})\bigg[2\sigma^2-\sigma^2({\eta}r+\acute{c})^2+\frac{\sigma^2}{\dot{\sigma}\eta^2}\bigg(\dot{\frac{A}{{\phi}X}\xi^\phi}\bigg)\bigg]}
\\\nonumber&+\frac{{Z^\phi_1}'}{4\eta^2({\eta}r+\acute{c})\bigg[({\eta}r+\acute{c})+4\eta^2+\sigma^2({\eta}r+\acute{c})\bigg]}.
\\\nonumber
&\varphi_{3}^{\phi}=\frac{\bigg[\sigma^2\eta2\eta(\frac{({\eta}r+\acute{c})^2}{4\eta^2})\dot{(\frac{A}{{\phi}X}\xi^\phi)'}-2({\eta}r+\acute{c}){Z^\phi_1}'\bigg]}
{\dot{\sigma}\bigg[2\eta^2-\sigma^2({\eta}r+\acute{c})^2+\frac{\sigma^2}{2\dot{\sigma}}({\eta}r+\acute{c})^2\dot{(\frac{A}{{\phi}X}\xi^\phi)}\bigg]}
\\\nonumber&+\frac{\bigg[2\eta^2+\sigma^2({\eta}r+\acute{c})^2+2\eta({\eta}r+\acute{c}){Z^\phi_1}'\bigg]\bigg[\frac{({\eta}r+\acute{c})^2}{4\eta^2}\dot{(\frac{A}{{\phi}X}\xi^\phi)'}\bigg]}
{({\eta}r+\acute{c})\dot{\sigma}\bigg[2\eta^2-\sigma^2({\eta}r+\acute{c})^2+\frac{\sigma^2}{2\dot{\sigma}}({\eta}r+\acute{c})^2\dot{(\frac{A}{{\phi}X}\xi^\phi)}\bigg]}.
\end{align}
\section*{Appendix B}
The expressions $\lambda_{1}^{\phi}$, $\lambda_{2}^{\phi}$ and
$\lambda_{3}^{\phi}$ appearing in equations \eqref{a145},
\eqref{a147} and \eqref{a148} are illustrated as
\begin{align}\nonumber
&\lambda_{1}^{\phi}=-2{Y^\phi}''
+\frac{\dot{Y^\phi}^2}{(\sigma^2r\psi-t\frac{\psi^2}{\sigma}+\omega_o)^2}
+\frac{1}{\breve{R_o}{\psi}e^{\sigma^2\frac{r^2}{2}-\sigma{\psi}tr
+\sigma^2r\frac{\omega_o}{\psi}+\frac{t^2}{2}\psi^2-\sigma_ot+Y^\phi}}.
\\\nonumber
&\lambda_{2}^{\phi}=\bigg[{\psi}r+t\frac{{\psi}^2}{\sigma}-\omega_o+\frac{\dot{Y^\phi}^2}{\sigma}\bigg]^2-
2\frac{\ddot{Y^\phi}}{({\psi}r-t\frac{\psi^2}{\sigma}+\omega_0)^2}
+2\frac{\dot{Y^\phi}}{\sigma^2\bigg({\psi}r-
t\frac{\psi^2}{\sigma}+\omega_0\bigg)^3}
{\psi}^2\\\nonumber&+\frac{\dot{Y^\phi}^2}{\sigma^2}\bigg[{\psi}r
-\frac{{\psi}^2}{\sigma}t+\omega_o\bigg]^2 +2\frac{{\psi}}{{\psi}r
-t\frac{{\psi}^2}{\sigma}+\omega_0}
\bigg[\sigma^2r-\sigma{\psi}t+\sigma^2\frac{\omega_o}
{{\psi}}+{Y^\phi}'\bigg]+\\\nonumber&\bigg(\sigma^2r-\sigma{\psi}t
+\frac{\omega_o}{{\psi}}\sigma^2
+{Y^\phi}'\bigg)-\frac{1}{\breve{R_o}{\psi}e^{\sigma^2\frac{r^2}{2}-\sigma{\psi}tr
+\sigma^2r\frac{\omega_o}{{\psi}}+\frac{t^2}{2}{\psi}^2-\sigma_ot+Y^\phi}}.
\\\nonumber
&\lambda_{3}^{\phi}=-\frac{\sigma^2}{{\psi}r-
t\frac{{\psi}^2}{\sigma}+\omega_0}
\bigg[{\psi}r+t\frac{{\psi}^2}{\sigma}-\omega_o+\frac{\dot{Y^\phi}^2}{\sigma}\bigg]^2
+\frac{\psi}{{\psi}r- t\frac{{\psi}^2}{\sigma}+\omega_0}+{Y^\phi}''
\\\nonumber&\bigg[\sigma^2r-\sigma{\psi}t+\sigma^2\frac{\omega_o}{{\psi}}
+{Y^\phi}'\bigg]+\bigg(\sigma^2r-\sigma{\psi}t+\frac{\omega_o}{{\psi}}\sigma^2
+{Y^\phi}'\bigg)-2\frac{\ddot{Y^\phi}}{{\psi}r-
t\frac{{\psi}^2}{\sigma}+\omega_0}\\\nonumber&+\bigg(\sigma^2r-\sigma{\psi}t+\frac{\omega_o}{{\psi}}\sigma^2
+{Y^\phi}'\bigg)^2.
\end{align}

\end{document}